\documentclass[11pt,a4paper]{article}
\usepackage{amsmath,amsfonts,amssymb,amsbsy}
\usepackage{mathrsfs,latexsym,tikz,slashed}
\usepackage{graphicx}
\usepackage{color}
\usepackage{booktabs}
\usepackage{multirow}
\usepackage{multicol}
\usepackage{subfig}
\usepackage{alpha}
\usepackage{mathrsfs}

\usepackage{macros_alpha}
\usepackage{textcomp}

\begin{document}

\renewcommand{\L}{\mathscr{L}}
\newcommand{\dlatt}[2][d]{\delta #2^{(#1)}}
\def\lcf{\Phi}

\newcommand{\EOMgf}{EOMs$|_\mathrm{gf}$}

\newcommand{\Green}[2][\mathrm{R}]{G\ifx \EMPTY#1\EMPTY \else_{#1}\fi[#2]}
\newcommand{\GreenOp}[4][\mathrm{R}]{G_{#3\ifx \EMPTY#1\EMPTY \else;#1\fi}^{(#4)}\left[#2\right]}

\newcommand{\GreenAmp}[2][\mathrm{R}]{\mathcal{G}\ifx \EMPTY#1\EMPTY \else_{#1}\fi[#2]}
\newcommand{\GreenAmpOp}[4][\mathrm{R}]{\mathcal{G}_{#3\ifx \EMPTY#1\EMPTY \else;#1\fi}^{(#4)}\left[#2\right]}
\newcommand{\GreenAmpOpT}[4][\mathrm{R}]{\slashed{\mathcal{G}}_{#3\ifx \EMPTY#1\EMPTY \else;#1\fi}^{(#4)}\left[#2\right]}

\newcommand{\Leff}{\L_\text{eff}}
\newcommand{\Llatt}{\L_\text{latt}}
\newcommand{\LeffG}{\L_\text{eff,G}}
\newcommand{\LlattG}{\L_\text{latt,G}}
\newcommand{\Lghosts}{\L_\mathrm{ghosts}}
\newcommand{\Lgf}{\L_\mathrm{gf}}
\newcommand{\Z}{\mathcal{Z}}

\newcommand{\A}{\mathcal{A}}

\newcommand{\eff}[2][\mathrm{R}]{#2_{\mathrm{eff}\ifx \EMPTY#1\EMPTY \else ;#1\fi}}

\newcommand{\define}{\stackrel{\text{def}}{=}}
\newcommand{\const}{\text{const.}}
\newcommand{\DR}{\mu^{4-D}}
\newcommand{\CA}{C_\text{A}}
\newcommand{\TF}{T_\text{F}}

\newcommand{\gammaE}{\gamma_\mathrm{E}}

\newcommand{\M}[1]{#1^{\mathrm{M}}}

\newcommand{\ord}{\mathrm{O}}
\newcommand{\PO}{\mathop{\mathrm{P}}}

\def\EMPTY{-EMPTY-}

\newcommand{\pint}[1]{\int\limits_{#1}}
\newcommand{\fint}[1]{\int\limits_0^1\D{x}}
\newcommand{\pD}{\mathcal{D}}
\newcommand{\D}{\text{d}}

\newcommand{\PI}[1][=]{\stackrel{\text{\tiny IBP}}{#1}}
\newcommand{\EOM}[1][=]{\stackrel{\text{\tiny EOM}}{#1}}
\newcommand{\BI}[1][=]{\stackrel{\text{\tiny BI}}{#1}}

\newcommand{\require}{\stackrel{!}{=}}

\renewcommand{\dim}[1]{\left[#1\right]}
\newcommand{\comm}[2]{\left[#1,#2\right]}
\newcommand{\acomm}[2]{\left\lbrace #1,#2\right\rbrace}
\renewcommand{\Re}{\mathop{\mathrm{Re}}}
\newcommand{\fderivative}[1]{\frac{\delta}{\delta #1}}
\newcommand{\Pexp}{\mathop{\mathrm{Pexp}}}

\newcommand{\Dferm}{\not\hspace{-3px}\mathfrak{D}}

\newcommand{\op}{\mathcal{O}}
\newcommand{\opT}{\slashed{\op}}
\newcommand{\Od}[2][]{
	\ifthenelse{ \equal {#1} {}}
	{\op^{(#2)}}
	{\op^{#1(#2)}}
}
\newcommand{\OdT}[2][]{\opT^{\ifx \EMPTY#1\EMPTY \else #1\fi (#2)}}

\newcommand{\base}{\mathcal{B}}
\newcommand{\based}[2][]{
	\ifthenelse{ \equal {#1} {}}
	{\base^{(#2)}}
	{\base^{#1(#2)}}
}

\newcommand{\R}[2][]{{#2}_{\ifx \EMPTY#1\EMPTY \else #1;\fi\mathrm{R}}}
\newcommand{\RGI}[2][]{{#2}_{#1\mathrm{RGI}}}
\newcommand{\bare}[2][]{{#2}_{\ifx \EMPTY#1\EMPTY \else #1;\fi 0}}
\newcommand{\lattExpCont}[1]{\Exp{#1}_{\mathrm{latt}(0);\mathrm{R}}}
\newcommand{\lattExp}[2][\mathrm{R}]{\Exp{#2}_{\mathrm{latt}(a)\ifx \EMPTY#1\EMPTY \else ;#1\fi}}
\newcommand{\contExp}[2][\mathrm{R}]{\Exp{#2}_{\mathrm{cont}\ifx \EMPTY#1\EMPTY \else ;#1\fi}}
\newcommand{\latt}[2][\mathrm{R}]{#2_{\mathrm{latt}(a)\ifx \EMPTY#1\EMPTY \else ;#1\fi}}
\newcommand{\cont}[2][\mathrm{R}]{#2_{\mathrm{cont}\ifx \EMPTY#1\EMPTY \else ;#1\fi}}

\def\lattExpRGI{\lattExp[\mathrm{RGI}]}
\def\contExpRGI{\contExp[\mathrm{RGI}]}

\newcommand{\ph}{\,{\scriptstyle\circ}\,}

\newcommand{\cev}[1]{\overset{\leftarrow}{#1}}
\renewcommand{\vec}[1]{\overset{\rightarrow}{#1}}

\newcommand{\ginfo}[2]{\substack{\displaystyle #1\\\displaystyle #2}}

\newcounter{opThree}
\setcounter{opThree}{-1}

\newcounter{opFour}
\setcounter{opFour}{-1}

\newcounter{opFive}
\setcounter{opFive}{-1}

\newcounter{opSix}
\setcounter{opSix}{-1}

\newcounter{baseFive}
\setcounter{baseFive}{-1}

\newcounter{baseSix}
\setcounter{baseSix}{-1}

\newcounter{baseGSix}
\setcounter{baseGSix}{-1}

\newcounter{opFiveT}
\setcounter{opFiveT}{-1}

\newcounter{opSixT}
\setcounter{opSixT}{-1}

\makeatletter
\newcommand{\ltxlabel}{\ltx@label}
\makeatother

\newenvironment{opThreeDef}[2][]{
\refstepcounter{opThree}\ltxlabel{#2}
\Od[#1]{-1}_{\theopThree}
}{}
\newcommand{\opThree}[2][]{\Od[#1]{-1}_{\ref{#2}}}

\newenvironment{opFourDef}[2][]{
\refstepcounter{opFour}\ltxlabel{#2}
\Od[#1]{0}_{\theopFour}
}{}
\newcommand{\opFour}[2][]{\Od[#1]{0}_{\ref{#2}}}

\newenvironment{opFiveDef}[2][]{
\refstepcounter{opFive}\ltxlabel{#2}
\Od[#1]{1}_{\theopFive}
}{}
\newcommand{\opFive}[2][]{\Od[#1]{1}_{\ref{#2}}}

\newenvironment{opSixDef}[2][]{
\refstepcounter{opSix}\ltxlabel{#2}
\Od[#1]{2}_{\theopSix}
}{}
\newcommand{\opSix}[2][]{\Od[#1]{2}_{\ref{#2}}}

\newenvironment{baseFiveDef}[2][]{
\refstepcounter{baseFive}\ltxlabel{#2}
\based[#1]{1}_{\thebaseFive}
}{}
\newcommand{\baseFive}[2][]{\based[#1]{1}_{\ref{#2}}}

\newenvironment{baseSixDef}[2][]{
\refstepcounter{baseSix}\ltxlabel{#2}
\based[#1]{2}_{\thebaseSix}
}{}
\newcommand{\baseSix}[2][]{\based[#1]{2}_{\ref{#2}}}

\newenvironment{baseGSixDef}[2][]{
\refstepcounter{baseGSix}\ltxlabel{#2}
\based[#1]{2}_{\thebaseGSix}
}{}
\newcommand{\baseGSix}[2][]{\based[#1]{2}_{\ref{#2}}}

\newenvironment{opFiveTDef}[2][]{
\refstepcounter{opFiveT}\ltxlabel{#2}
\OdT[#1]{1}_{\theopFiveT}
}{}
\newcommand{\opFiveT}[2][]{\OdT[#1]{1}_{\ref{#2}}}

\newenvironment{opSixTDef}[2][]{
\refstepcounter{opSixT}\ltxlabel{#2}
\OdT[#1]{2}_{\theopSixT}
}{}
\newcommand{\opSixT}[2][]{\OdT[#1]{2}_{\ref{#2}}}

\def\FORM/{\texttt{FORM}}
\def\QGRAF/{\texttt{QGRAF}}

\newcommand{\finite}{\text{finite}}

\newcommand{\splitEq}[2]{
\\[-5pt]\begin{minipage}{0.5\textwidth}
\begin{align*}
#1
\end{align*}
\end{minipage}
\begin{minipage}{0.5\textwidth}
\begin{align}
#2
\end{align}
\end{minipage}
\ \\}

\makeatletter
\newsavebox{\@brx}
\newcommand{\llangle}[1][]{\savebox{\@brx}{\(\m@th{#1\langle}\)}
  \mathopen{\copy\@brx\mkern2mu\kern-0.9\wd\@brx\usebox{\@brx}}}
\newcommand{\rrangle}[1][]{\savebox{\@brx}{\(\m@th{#1\rangle}\)}
  \mathclose{\copy\@brx\mkern2mu\kern-0.9\wd\@brx\usebox{\@brx}}}
\makeatother

\newcommand{\UN}[2][N]{\ensuremath{\mathrm{U}(#1)\ifx \EMPTY#2\EMPTY \else _\mathrm{#2}\fi}}
\newcommand{\SUN}[2][N]{\ensuremath{\mathrm{SU}(#1)\ifx \EMPTY#2\EMPTY \else _\mathrm{#2}\fi}}
\newcommand{\Times}{\mathop{\times}}

\newcommand{\C}{\mathcal{C}}
\renewcommand{\P}{\mathcal{P}}
\newcommand{\T}{\mathcal{T}}

 \newcommand{\obs}{\mathcal{P}}
\newcommand{\ainv}{a^{-1}}
\newcommand{\con}{\mathrm{con}}
\newcommand{\Slat}{S_\lat}
\newcommand{\contrs}{\mathrm{cont}}
\newcommand{\Scont}{S_\contrs}
\newcommand{\Lcont}{\L_\contrs}
\newcommand{\Lfp}{\L_\mathrm{fp}}
\newcommand{\Lbf}{\L_\mathrm{bf}}
\newcommand{\nmin}{{n_\mathrm{min}}}

\newcommand{\ranglat}{\rangle_\lat}
\newcommand{\rangcont}{\rangle_\contrs}
\newcommand{\ranglatcon}{\rangle_\lat^\con}
\newcommand{\rangcontcon}{\rangle_\contrs^\con}

\newcommand{\dPhi}{\delta \Phi}
\newcommand{\mel}{\mathcal{M}}
\newcommand{\melr}{\mathcal{M}^\mathrm{R}}
\newcommand{\melrgi}{\mathcal{M}^\mathrm{RGI}}
\newcommand{\opr}{\op^\mathrm{R}}
\newcommand{\opE}{\mathcal{E}}
\newcommand{\Zoffd}{A}

\newcommand{\basergi}{\base^\mathrm{RGI}}
\newcommand{\baser}{\base^\mathrm{R}}

\newcommand{\gammahat}{\hat\gamma}

\newcommand{\dC}{\delta C}

\newcommand{\Cik}{C_{ik}}
\newcommand{\CikO}{C^\op_{ik}}
\newcommand{\CjkO}{C^\op_{jk}}
\newcommand{\Cred}{C^{\mathrm{red}}}
\newcommand{\ClkOred}{\Cred_{lk}}

\newcommand{\Zoffdbar}{\bar \Zoffd}
\newcommand{\Zbar}{\bar Z}

\newcommand{\opferm}{\op^\mathrm{P}}

\newcommand{\lambdar}{\lambda_{{{\rm R}}}} 
\preprintno{
DESY 19-188
}

\title{Asymptotic behavior of cutoff effects in
  Yang-Mills theory and in Wilson's lattice QCD
}

\author[desy1]{Nikolai~Husung}
\author[desy2]{Peter~Marquard}
\author[desy1,hu]{Rainer~Sommer}
\address[hu]{Institut~f\"ur~Physik, Humboldt-Universit\"at~zu~Berlin,
             Newtonstr.~15, 12489~Berlin, Germany}
\address[desy1]{John von Neumann Institute for Computing (NIC),
               DESY, Platanenallee~6, D-15738 Zeuthen, Germany}
\address[desy2]{Theory group, DESY, Platanenallee~6, D-15738 Zeuthen, Germany}

\vspace*{-1cm}

\begin{abstract}
Discretization effects of lattice QCD are described by 
Symanzik's effective theory when the lattice spacing, $a$, 
is small. Asymptotic freedom predicts that the leading 
asymptotic behavior is 
$\sim  a^\nmin   [\gbar^2(a^{-1})]^{\gammahat_1} \sim 
 a^\nmin   \left[\frac{1}{-\log(a\Lambda)}\right]^{\gammahat_1}$. 
For spectral quantities, 
$\nmin=d$ is given
in terms of the (lowest) canonical dimension, $d+4$, of the 
operators in the local effective Lagrangian and 
$\gammahat_1$ is proportional to the leading eigenvalue 
of their one-loop anomalous dimension matrix $\gamma^{(0)}$. 
We determine $\gamma^{(0)}$ for Yang-Mills theory ($\nmin=2$) and
discuss consequences in general and  for  perturbatively 
improved short distance observables. With the help of 
results from
the literature, we also discuss the $\nmin=1$ case of 
Wilson fermions with perturbative $\ord(a)$ improvement
and the discretization effects specific to the flavor 
currents. In all cases known so far, the discretization effects
are found to disappear faster than the naive $\sim a^\nmin$
and the log-corrections are a rather weak 
modification -- in contrast to the two-dimensional O(3) sigma model.
\end{abstract}


\maketitle 
 
\tableofcontents

\section{Introduction}

Lattice regularizations provide a definition of 
quantum field theories beyond 
perturbation theory. Evaluating the associated path integral by Monte Carlo  also constitutes a non-perturbative 
calculational method to derive predictions from the theory. 
One of the systematic effects that have to be taken into 
account is the dependence of results on the lattice spacing $a$
(we assume a hyper-cubic lattice throughout) or in other words 
the size of discretization errors,
\begin{equation} \label{eq:cutoffeffect}
  \Delta_\obs(a)  = \obs(a) - \obs(0)\,, 
\end{equation}
associated with a dimensionless observable 
$\obs$ of the theory. As a start, one may consider 
the classical field theory. One then has smooth fields,
and the lattice-Lagrangian can simply be Taylor expanded.
It is the continuum one 
up to terms suppressed by powers of~$a$. 

One may therefore think that also in the full, quantized, theory 
the small-$a$ behavior of 
the discretization errors is $\Delta_\obs(a) = p_1 a^{\nmin} +
p_2 a^{\nmin+1} + \ldots $ with the integer $\nmin$ 
given by the first non-zero power in the classical 
Taylor expansion of the Lagrangian. However, the divergences of quantum field theories 
spoil this behavior.

Still, precise statements can be made 
about the small-$a$ expansion, based on Symanzik's effective
theory (SymEFT) ~\cite{Symanzik:1979ph,Symanzik:1981hc,Symanzik:1983dc,Symanzik:1983gh}, see also~\cite[p.~39ff.]{Weisz:2010nr}.
It describes the small-$a$ behavior by an effective field theory
with a local Lagrangian 
\begin{equation}\label{eq:effLagrangian}
\Leff(x)=\L+a\dlatt[1]{\L}(x)+a^2\dlatt[2]{\L}(x)+\ldots\,.
\end{equation}
The effective theory can be thought of as a continuum effective
theory, regularized e.g. by dimensional regularization. The
first term is the continuum Lagrangian $\L$ of the fundamental 
field theory and $\dlatt[d]{\L}(x)$ are local Lagrangians
of higher mass dimension. The leading term in \eq{eq:cutoffeffect} is then
given by the one\footnote{We will be more precise below.}with the lowest mass dimension in 
\eq{eq:effLagrangian}, i.e. $\dlatt[1]{\L}(x)$, unless it vanishes.
The corrections $\dlatt[d]{\L}(x)$ can be written as a linear
combination of basis operators $\base_i(x)$ with the appropriate canonical mass
dimensions. Renormalization of the effective theory introduces 
anomalous dimensions for the operators $\base_i$. It may therefore 
modify the small-$a$ expansion to
$\Delta_\obs(a) = p_1 a^{\nmin + \eta} + \ldots $
with, in general, non-integer $\eta$. The (leading) anomalous dimension $\eta$
is in general a non-perturbative quantity, but
it may sometimes be estimated by perturbation theory
in the $\epsilon$-expansion, see \cite{ZinnJustin:2002ru}.

We now turn to asymptotically free theories such as QCD. 
There, small $a$ 
means weak coupling at the scale of the lattice cutoff 
and the anomalous dimension
can 1) be computed in perturbation theory and 2) it leads to
a modification of $a^n$ by logs \cite{Symanzik:1979ph,Symanzik:1981hc,Balog:2009yj,Balog:2009np}, \begin{equation}
   \Delta_\obs(a) = p_1 [-\log(a\Lambda)]^{-\hat{\gamma}}  \,a^{\nmin} + \ldots 
   \label{eq:logcorr}
\end{equation}
and not by fractional powers. The intrinsic scale of the theory, 
$\Lambda$, is a 
renormalization group invariant and the 
exponent $\hat{\gamma}$ is proportional to a one-loop anomalous dimension.
Since the work of \cite{Luscher:1991wu}, continuum extrapolations are routinely 
performed in order to obtain quantitative numbers for continuum
field theory observables. They have been carried out with 
just powers\footnote{Sometimes an additional power of $\gbar^2(\ainv) \sim [-\log(a\Lambda)]^{-1}$ has been used when a tree-level improved action is used. Here $\gbar^2$ is the running coupling in some
scheme.}of $a$, thus implicitly {\em assuming 
 that $\hat{\gamma}$ is small}.
Of course this can not really be taken for granted until $\hat{\gamma}$ 
is known from a computation. We here start to fill this gap.

Note that the logarithmic corrections in \eq{eq:logcorr} 
can be very relevant. An explicit example is provided by the seminal work of Balog, Niedermayer and Weisz \cite{Balog:2009yj,Balog:2009np}.
It concerns the 2-d O(3) sigma model where the leading term is
$\hat{\gamma} =-3$ and the logarithmic corrections change the
naive $a^2$ behavior to a shape which numerically looks like
$a$ in a broad range of $a\Lambda$ \cite{Balog:2009yj,Balog:2009np}. This numerical behavior
led to quite some concern \cite{Knechtli:2005jh} and the computation of the 
logarithmic corrections by 
Balog, Niedermayer and Weisz were essential to confirm that 
the SymEFT description holds and put continuum extrapolations 
on a solid ground. In lattice QCD, knowledge of the leading 
power of the logarithms (and partially awareness of the issue) are still missing;
in particular it is important to have a 
confirmation that $\hat\gamma$ is small as is usually assumed. Let us cite Peter
Weisz \cite{Weisz:2010nr}:
\\[1ex]
{\it The program should be carried out for lattice actions used for 
large scale simulations of QCD, when technically possible, in order 
to check if potentially large logarithmic corrections to lattice 
artifacts predicted by perturbative analysis appear.
}
\\[1ex]
Ten years later, as a first step, we do carry out the program 
in the pure Yang-Mills (YM) theory
as well as in Wilson's  lattice QCD  without 
non-perturbative $\ord(a)$
improvement. The latter case is rather simple and basically
given by results in the literature. We will therefore discuss
only the YM theory in detail and just mention the difference and results in Wilson's QCD in \sect{s:Wils}.

\subsection*{Scope}\addcontentsline{toc}{subsection}{Scope}

In addition to the discretization effects due to the
terms $\dlatt[d]{\L}$ in the effective Lagrangian, 
correlation functions of local fields $\Phi(x)$ 
also get $a$-effects from corrections to the fields
$\Phi(x)$ represented 
in the SymEFT  \cite{Heatlie:1990kg, Luscher:1996sc}.
Apart from mostly restricting ourselves
to the YM theory, we also do not discuss these additional discretization effects.
They are absent in quantities which are
independent of details of the local fields. We call those spectral quantities,
since the spectrum of the Hamiltonian is the important application.
In the YM theory, correlation functions themselves have 
so far not played a relevant role, apart from one notable exception.
The exception is the new sector of Gradient flow observables \cite{Narayanan:2006rf,Luscher:2010iy}. We leave its treatment to future work.
 
\section{Symanzik effective theory and logarithmic
corrections to $a^n$ behavior}
\label{s:sym}

We consider YM theory in 4 dimensions defined by the action
\bes
  S_\mathrm{lat} &=& \frac2{g_0^2}\, \sum_{x,\mu>\nu=0}^3\,  p(x,\mu,\nu)\,, 
  \nonumber \\[-1ex]\label{eq:Slat}\\[-1ex] \nonumber
  && p(x,\mu,\nu)=\Re\,\tr\,(1-U(x,\mu)U(x+a\hat\mu,\nu)U^{-1}(x+a\hat\nu)
                   U^{-1}(x,\nu) )
\ees
in terms of the link variables $U(x,\mu) \in $~SU(N), connecting $x+a\hat\mu$ and $x$. We assume a lattice with periodic boundary conditions in space and infinite (or arbitrarily large) time extent.\footnote{
In practice, finite lattices are of course needed for the
Monte Carlo evaluation. The appropriate modifications of 
equations such as \eq{eq:mass} are standard.}

As an example of a simple observable, $\obs$, take a ratio of glue-ball masses, which may be defined as ($\partial_\mu^\mathrm{lat}  f(x) =[f(x+a\hat\mu)-f(x)]/a$ and $x=(x_0,\vecx)$)
\bes
  \label{eq:mass}
  \obs=\mh_i/\mh_j,\quad
  \mh_i = -\lim_{x_0\to\infty} 
  \partial_{0}^\mathrm{lat} \log\left(   a^3\sum_\vecx C_i(x) \right) \,, 
\ees
in terms of a two-point function
\bes
  \label{eq:Cx}
  C_i(x-y) &=& \langle\,  \Phi_i(x) \Phi_i(y) \,\ranglatcon 
\ees
The gauge invariant fields $\Phi_i(x)$ are formed out of small (with a maximal extent
$r_\mathrm{w}$ with $r_\mathrm{w}/a$ fixed) spatial Wilson loops,
combined in such a way as to have a definite transformation under the 
lattice cubic group. A very simple example is
the scalar field $\Phi_1(x)=Z_{F^2}\sum_{k,l\in\{1,2,3\}}p(x,k,l)$. For simplicity we assume in the following
that the renormalization factors, such as $Z_{F^2}$ are
determined such that they do not introduce any cutoff effects.
In perturbation theory minimal (lattice) subtraction has this
property. 
Expectation values are defined by the lattice path integral
\bes
    \label{eq:obslat}
   \langle F(U) \ranglat = \frac1{\cal Z}\int \prod_{x,\mu}\rmd U(x,\mu) 
   \rme^{-\Slat(U)} F(U) \,, \quad  
\ees
where $\cal Z$ normalizes such that $\langle 1 \ranglat=1$, $F(U)$ stands for a function of any number of link variables $U(x,\mu)$ and $\rmd U(x,\mu)$ is the 
invariant Haar measure. The label ``con'' stands for connected
correlation functions, namely the subtraction of 
$[\langle\,  \Phi_i(x)\,\ranglat]^2$ in \eq{eq:Cx}.

Note that while $C_i(x)$ depend on the details of the definition of 
$\Phi_i(x)$,  the masses $\mh_i$ only depend on the quantum numbers of the field 
$\Phi(x)$. Masses or more generally energies are spectral quantities.

SymEFT gives the small-$a$ expansion of correlation functions
such as $C_i(x)$ in the form of a continuum effective field theory. 
The central statement is 
\bes \label{eq:Cxexp}
  C(x) = C^\contrs(x) + a^\nmin \dC(x) + \ord(a^{\nmin+1}) 
\ees
and the expansion on the r.h.s. can be obtained from 
the effective continuum field theory with effective Lagrangian 
\eq{eq:effLagrangian} supplemented by correction terms
which are due to correction terms 
of the fields \cite{Heatlie:1990kg,Luscher:1996sc}
\begin{equation}\label{eq:discretisedQuantity}
\eff[]{\lcf}(x)=\lcf(x)+a\dlatt[1]{\lcf}(x)+a^2\dlatt[2]{\lcf}(x)+\ldots\,.
\end{equation}
Let us mention right away that $\nmin=2$ in the considered YM theory.

For precise statements we need to specify 
\begin{enumerate}
  \item the rules of the EFT, i.e. how precisely are $\dC(x)$ defined in 
  terms
  of $\dlatt{\L}(x), \dlatt{\lcf}(x)$,
  \label{it:rules}
  \item which local operators contribute to $\dlatt{\L}(x)$, $\dlatt{\lcf}(x)$,
  \label{it:fields}  
  \item how are the parameters of the EFT determined, in other words 
  how are the coefficients of those operators contributing to $\dlatt{\L}(x), \dlatt{\lcf}(x)$ determined.
  \label{it:match}  
\end{enumerate}
We discuss these items in turn.

\noindent {\bf \ref{it:rules}.} 
The correction terms $\dlatt{\L}(x)$
etc. have canonical mass dimension $4+d$. A path integral with weight
$\rme^{-\int\rmd^4x \Leff(x)}$ is thus not renormalizable. Path integral expectation values  are {\em defined} by expanding in the parameter $a$ before 
integrating over the fields. For our example, \eq{eq:Cxexp}, we then have as 
a definition of $\dC(x)$
\bes
\label{eq:C1eff}
    \dC(x) &=& \dC^\L(x) + \dC^\Phi(x) \,,
    \\ 
\label{eq:C1S}
    \dC^\L(x-y) &=&-\int\rmd^4 z \, \langle\,  \Phi(x) \Phi(y)\, \dlatt[2]{\L}(z)\,\rangcontcon 
  \\ 
\label{eq:C1Phi}
 \dC^\Phi(x-y)  &=&\langle\,  \dPhi(x) \Phi(y)\,\rangcontcon 
  + \langle\,  \Phi(x) \dPhi(y)\,\rangcontcon 
\ees
where $\langle\,  X \,\rangcontcon$ is given by the standard continuum connected correlation function 
with continuum Lagrangian 
\bes
\label{eq:scont}
\Lcont(A) = -\frac{1}{2g_0^2}\,\sum_{\mu,\nu}\tr(F_{\mu\nu}(A) F_{\mu\nu}(A))\,, \quad 
F_{\mu\nu}(A) = [D_\mu(A),D_\nu(A)]\,,
\ees
written in terms of the covariant derivative 
\bes
\label{eq:Dmu}
D_\mu(A) = \partial_\mu + A_\mu \,.
\ees
We have already anticipated that \ref{it:fields}. leads to the vanishing of $\dlatt[1]{\L}, \dlatt[1]{\lcf}$
and used a shorthand $\dPhi=\dlatt[2]{\lcf}$.

\noindent {\bf \ref{it:fields}.}
The correction Lagrangians $\dlatt{\L}$ are linear combinations 
\bes
  \dlatt{\L}(x)= \sum_i \omegasym_i(g_0^2) \,\op_i(x)  
\ees
of 
local operators $\op_i(x) $ which comply with the symmetries of the underlying lattice 
theory and have a mass dimension $4+d$.  Gauge invariance is one of the symmetries (gauge fixing is needed only in 
\sect{s:AD} where we report on the perturbative computation). 
One may further drop all
combinations of fields which vanish by the continuum equation of motion,
$[D_\mu, F_{\mu\nu}(x)] =0$,
(such as $\op=\tr([D_\mu, F_{\mu\nu}]\,[D_\rho F_{\rho\nu}])$)
\cite{Luscher:1996sc} as well as all operators which 
can be written as total derivatives of the
form $\slashed{\op} = \partial_\mu K_\mu$. 
After doing that, we have a so 
called  ``on-shell'' basis. For YM it consists of
two operators, which we may choose as
\bes
\op_{1}=\frac{1}{g_0^2}\sum\limits_{\mu,\nu,\rho}\tr([D_\mu, F_{\nu\rho}]\,[D_\mu, F_{\nu\rho}])
\,,\quad
\op_{2}=\frac{1}{g_0^2}\sum\limits_{\mu,\nu}\tr([D_\mu, F_{\mu\nu}]\,[D_\mu, F_{\mu\nu}])\,,
\label{eq:ops}
\ees
already known from Refs.~\cite{Weisz:1982zw,Luscher:1985}\footnote{That
reference discusses the construction of a lattice improved action
such that the $a^2$ terms in the SymEFT are absent. The basis of operators
is the same.}. Note that
$\op_{2}$ breaks the O(4) rotational invariance of the continuum 
Lagrangian \eq{eq:scont} down to 90$^{\circ}$ rotations around the lattice axes. 
Dropping it, one has the general effective Lagrangian of a low energy theory
with just gauge fields and O(4) invariance. 
This is a (tiny) sector of the Lagrangian considered for beyond the standard model 
phenomenology in Ref.~\cite{Alonso:2013hga}. 
The operator, $\frac{1}{g_0^3}\tr(F_{\mu\nu}F_{\nu\rho}F_{\rho\mu})$, considered there is seen to be on-shell equivalent to 
\bes
\op_1=\frac{2}{g_0^2}\sum\limits_{\mu,\nu,\rho}\left(\tr([D_\mu,F_{\mu\nu}][D_\rho,F_{\rho\nu}])-\tr(F_{\mu\nu}F_{\nu\rho}F_{\rho\mu})\right)+\text{(total divergences)}
\ees 
using integration by parts and the Bianchi identity.
Gauge invariant dimension five operators do not exist and thus 
YM theory has $\nmin=2$. The corrections to the continuum fields $\Phi_i$ will not be needed. 

Now we consider the $a$ expansion of our observable,
\bes
  \obs = \obs^\contrs + 
               a^2 [\delta\obs^{\L} + \delta\obs^{\Phi}] + \ord(a^3) \,.
\ees  
Inserting the spectral representations into the ratios 
$C^\L_i/C^\contrs_i$ which appear as one expands the r.h.s. of
\eq{eq:mass} in $a$, one sees\footnote{For intermediate steps in the derivation, see \cite{Sommer:2010ic}, sect.~9.4.1. In quantum mechanics the relation given is the Feynman-Hellmann theorem.} 
\bes
  \label{eq:S2matel}
  \delta\obs^\L = - \frac12\,
  [\langle i| \dlatt[2]{\L}(0) | i\rangle\,
  -\langle j| \dlatt[2]{\L}(0) | j \rangle]\,, \quad
  \delta\obs^\mathrm{\Phi} = 0\,.
\ees
The states $|i\rangle$ with $\langle i|i\rangle =2L^3$ are the 
 ground states of the Hamiltonian of the finite volume theory with spatial volume $L^3$ in the zero momentum sector
of the Hilbert space with the quantum numbers of $\Phi_i$. 
The vanishing of $\delta\obs^\mathrm{\Phi}$ was to be expected as the energy of a physical state 
should not depend on the interpolating field used to create it, including its renormalization.
Since physical quantities which do depend on $\dPhi$ have so far 
not been in the focus of lattice computations, and also because
each field appearing in the correlation functions
has to be considered separately, we will ignore the
contribution of $\dPhi$ from now on. We concentrate on spectral 
quantities.

\noindent {\bf \ref{it:match}.}
The coefficients $\omegasym_i$ are needed, in particular their dependence 
on the parameters of the theory. 
\Eq{eq:S2matel} makes it clear that actually we first have to
renormalize the operators $\op_i$ and then determine their coefficients
by matching, which will be discussed in \sect{s:match}. Renormalization  
introduces a dependence on the renormalization scale $\mu$ (and scheme). By renormalization group improvement we turn it into a dependence on
the lattice spacing, which we are seeking. In the 2-d O(N) sigma model,
all this has been done to next-to-leading order in the coupling\cite{Balog:2009yj} .
Here we are content with the leading order since it 
predicts the asymptotic behavior of $\Delta_\obs$.

Before proceeding it is convenient to switch to a basis of operators, with elements 
$\base_i=\sum_jv_{ij}\op_j$ which do not mix at one-loop order, i.e.
\bes
   \baser_i(\mu) = [1+g^2 Z^{(1)}_i + \ord(g^4)]\, \base_i\,,
\ees 
where $\baser_i(\mu)$ denote the renormalized operators in some scheme at
renormalization scale~$\mu$. One may think of the $\msbar$ scheme.

In general, we then have 
$
   \Delta_\obs = \sum_i \csym_i \melr_{\obs,i}, 
$
where at leading order in the coupling
$\omega_j=\omegasym_j^{(n)}g_0^{2n} + \rmO(g_0^{2n+2})\,,\; \omegasym_j^{(n)}=\sum_i \csym_i^{(n)} v_{ij} $ and
$\melr_{\obs,i}$ are matrix elements of the operators $\base_i$ in the
continuum field theory. The renormalized matrix elements are denoted
\bes  \label{eq:melR}
   \melr_{\obs,i}(\mu) = \langle \psi_\obs| \baser_i(\mu) |\psi_\obs\rangle \,,
\ees
with some physical state $|\psi_\obs\rangle$, analogous to $|\psi_\Phi\rangle$.
We have suppressed the spacetime argument of $\base_i$.

The coefficients $\csym_i$ depend on the renormalization scheme adopted for $\base_i^\mathrm{R}$ as well
as on $\mu$ and $a$. 
We may thus write 
(dropping higher powers of $a$ without notice)
\bes
   \label{eq:deltap1}
   \Delta_\obs(a) = -a^2 \sum_i \csym_i(\gbar(\mu),a\mu)\, \melr_{\obs,i}(\mu)\,,
\ees
where the dependence of $\csym_i$ on $\mu$ cancels the one of $\melr_{\obs,i}(\mu)$. 

In order to systematically learn about the behavior for small $a$ we use
renormalization group improvement, namely we set $\mu=1/a$, and
introduce the renormalization group invariant matrix elements
\bes
     \melrgi_{\obs,i}  = \sum_j \varphi_{ij}(\gbar(\mu))\, \melr_{\obs,j}(\mu) 
     = \langle \psi_\obs| \basergi_i |\psi_\obs\rangle \,.
\ees
The matrix valued function ($\Pexp$ denotes path ordering: terms with smallest 
$x$ appear to the left)
\bes
    \label{eq:phifct}
   \varphi(\gbar) &=&\left[\,2b_0 \gbar^2\,\right]^{-\gamma^{(0)}/2b_0}
         \Pexp\left\{-\int_0^{\gbar} \rmd x
                     \left[\,{ \gamma(x) \over\beta(x)}
                           -{\gamma^{(0)}\over b_0 x}\,\right]
                     \right\}\,,
      \\ &=& \left[\,2b_0 \gbar^2\,\right]^{-\gamma^{(0)}/2b_0} \times 
                     [1+\rmO(\gbar^2)]
                     \label{eq:phifct2}
\ees
involves the anomalous dimension matrix $\gamma$ defined by
\bes
    \mu \frac{\rmd}{\rmd\mu} \baser_i(\mu) = \sum_j \gamma_{ij}(\gbar(\mu))  \,\baser_j(\mu) \,.
\ees
It has the expansion
\bes
   \gamma(\gbar) = -\gbar^2 \,[\gamma^{(0)}  +\gamma^{(1)} \gbar^2 + \ldots ]\,,
\ees
where by our choice of basis $\gamma^{(0)}$ is diagonal, 
\bes 
\label{eq:gammahat}
        \frac1{2b_0}\gamma^{(0)} =\diag(\gammahat_1,\gammahat_2)\,.
\ees
Our convention for the $\beta$-function is 
$\beta(\gbar(\mu)) = \mu \frac{\rmd}{\rmd\mu} \gbar(\mu)$ with expansion
$\beta(\gbar)=-\gbar^3\,(b_0 +b_1 \gbar^2 +\ldots) $.

Asymptotic freedom means that perturbation theory is applicable at small $a$. 
The asymptotic behavior of \eq{eq:deltap1} can thus be inferred from 
(renormalized) perturbation theory. The
$\ord(g^2)$ term in \eq{eq:phifct2} is then subdominant and further we may expand
\bes
    \label{eq:ciexp}
    \csym_i(\gbar(a^{-1}),1) = \csym_i^{(0)} +  \csym_i^{(1)} \gbar^2(a^{-1}) + \ldots \,.
\ees
Putting everything together and concentrating on the leading term we arrive at 
\bes
   \Delta_\obs(a) &=& -a^2 \sum_{i} \csym_i^{(0)}  
   \left[\,2b_0 \gbar^2(a^{-1})\right]^{\gammahat_i}  \melrgi_{\obs,i}\,[1 +\ord(\gbar^2(a^{-1})]+\ord(a^4)\,.
\ees
Ordering $\gammahat_1 < \gammahat_2$, the leading asymptotics is
\bes
   \Delta_\obs(a) \sim  a^2   \left[\,2b_0 \gbar^2(a^{-1})\right]^{\gammahat_1} 
    \sim  a^2   \left[\frac{1}{-\log(a\Lambda)}\right]^{\gammahat_1} \,,
\ees
unless $\csym_1^{(0)}$ or the matrix element $\melrgi_{\obs,1}$ vanish. Generically, there is no
reason for the latter to do so. A positive/negative $\gammahat_1$ leads to 
an accelerated/decelerated asymptotic convergence as compared to naive $a^2$ behavior.

\section{One-loop computation of the anomalous dimension matrix}
\label{s:AD}
We now turn to the anomalous dimension matrix $\gamma^{(0)}$. 
Although the renormalization of composite pure gauge theory operators has been
discussed extensively \cite{Gracey:2002he,Alonso:2013hga}, a new computation 
is necessary because of the rotation symmetry violating operator $\op_2$, \eq{eq:ops}, which is not found in the literature. We thus employed dimensional regularization 
and computed the renormalization matrix,
\begin{equation}
\R{\begin{pmatrix}
    \op_1\\[3pt]
    \op_2
  \end{pmatrix}}=
\begin{pmatrix}
    Z_{11}& 0\\[3pt]
    Z_{21}&Z_{22}
\end{pmatrix}
\begin{pmatrix}
        \op_1\\[3pt]
        \op_2
    \end{pmatrix}\,,\label{eq:Zop}
\end{equation}
to one-loop order.
Here $Z_{12}$ vanishes because dimensional regularization
preserves rotational symmetry and thus $\R{(\op_1)}$ can 
not have a rotational non-invariant piece $Z_{12} \op_2$.

The $Z$-matrix is obtained from a perturbative 
computation of a sufficient number of 
expectation values 
\bes
  \CikO = \langle  \op_i \op_{k}^\mathrm{probe} \rangle 
\ees
of the operators $\op_i$ together with 
suitable multi-local, renormalized, operators $\op_{k}^\mathrm{probe}$. We may choose 
$\op_{k}^\mathrm{probe}$ including their kinematics to simplify the computation.
Unfortunately, just choosing  them 
to be composed of local gauge invariant operators, 
e.g. $\tr F_{\mu\nu}F_{\mu\nu}$,
one quickly discovers that one-loop computations 
are insufficient, since the tree-level 
correlation functions vanish. 

\begin{figure}\centering
\subfloat[Two-point function.]{\includegraphics[scale=1.3]{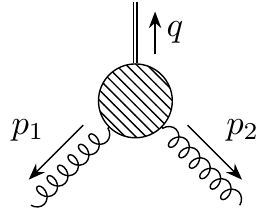}}\qquad
\subfloat[Three-point function.]{\includegraphics[scale=1.3]{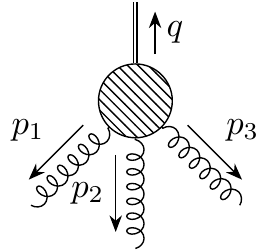}}
\caption{Schematic representation of the needed two-point and
three-point functions 
 with insertion of an operator $\op_i$. The "blob" represents all possible connected tree-level and one-loop graphs with given number of external legs.}
 \label{f:diagrams}
\end{figure}

As one option, we thus relaxed on manifest 
gauge invariance of $\Cik$ and
consider gauge dependent Green's functions 
with
\bes
    \op_{1}^\mathrm{probe} = \tilde{A}^a(p_1)\cdot\eta_1\;
    \tilde{A}^b(p_2)\cdot\eta_2\,,\quad
    \op_{2}^\mathrm{probe} = \tilde{A}^a(p_1)\cdot\eta_1\; \tilde{A}^b(p_2)\cdot\eta_2\; \tilde{A}^c(p_3)\cdot\eta_3\,, 
\ees
in terms of the momentum space fields
$\tilde A_\mu(p) =\int \rmd^4x\, \rme^{-ipx} A_\mu(x)$.
We have
$\sum_i p_i=-q$ as indicated in \fig{f:diagrams} and 
choose $[(p_i)_0]^2=-(\vecp_i)^2$, $p_i\cdot\eta_i=0$ for all $i$ with the Euclidean scalar product $p\cdot\eta=\sum_\mu p_\mu\eta_\mu$.
In principle  mixing of $\op_i$ 
with gauge-non-invariant operators then has to be taken into account \cite{Joglekar:1975nu,Collins:1994ee}. However, those do not contribute to the 
on-shell Green's functions selected by our choice of kinematics.  
Since we want to restrict ourselves to the 
two and three gluon $\op^\mathrm{probe}$
from above, we need to
have a non-zero momentum $q$ of the operators $\op_i$. Otherwise the Green's functions vanish by kinematics. The price to pay is that $\op_i$ mix with 
the ``total divergence operators'', 
\bes
   \slashed\op_1 = \frac{1}{g_0^2}\sum\limits_{\mu,\nu,\rho}\partial_\mu\tr(F_{\rho\nu}\, [D_\mu, F_{\rho\nu}])\,,
  \quad 
   \slashed\op_2 = 
   \frac{1}{g_0^2} \sum\limits_{\mu,\nu}\partial_\mu\tr(F_{\mu\nu}\,[D_\mu, F_{\mu\nu}])\,,
\ees
as 
\begin{equation}
\R{\begin{pmatrix}
    \op\\[3pt]
    \slashed\op
  \end{pmatrix}}=
\begin{pmatrix}
    Z& \Zoffd^{\op\slashed{\op}}\\[3pt]
    0&Z^{\slashed\op}
\end{pmatrix}
\begin{pmatrix}
        \op\\[3pt]
        \slashed\op
    \end{pmatrix}\,,\label{eq:Zopgaugevariant}
\end{equation}
with a block-triangular structure. 

As a second option, we considered the background field
method\cite{DeWitt:1967ub,KlubergStern:1974xv,Abbott:1980hw,Luscher:1995vs}. 
It consists of introducing a smooth classical background field, $B_\mu(x)$. The gauge field, 
\bes
   A_\mu= B_\mu + g_0 Q_\mu \,,
\ees
is split into the background field and the quantum fluctuations 
$Q_\mu$. Note that the background field is
{\em not} required to satisfy the equation of motion.
In addition to the Lagrangian
\bes
    \Lbf(B,q) = \Lcont(B+g_0Q)\,,
\ees
one chooses the background field gauge with gauge-fixing term
\bes
  \Lgf(B,Q) = -\lambda_0\,\sum_{\mu,\nu}\tr([D_\mu(B),Q_\mu] [D_\nu(B),Q_\nu] 
\ees
instead of the standard $-\lambda_0\,\tr((\partial_\mu A_\mu) (\partial_\nu A_\nu)) $ and adds a Faddeev Popov term \cite{Faddeev:1967}.

In this case, we can form 
\bes
    \op_{1}^\mathrm{probe} = \tilde{B}^a_\mu(p_1)\;
    \tilde{B}^b_\nu(p_2)\,,\quad
    \op_{2}^\mathrm{probe} = \tilde{B}^a_\mu(p_1)\; \tilde{B}^b_\nu(p_2)\; \tilde{B}^c_\rho(p_3)\,, 
\ees
just in terms of the background field,
and obtain gauge invariant
$\Cik$ by construction. We can remain with Euclidean momenta
and do not need a nonzero momentum to flow into the operator $\op_i$. 
Thus the mixing with total divergence operators does not contribute any more.
The downside is that here the equations of motion
do not hold. Therefore, we have to consider the 
mixing structure
\begin{equation}
\R{\begin{pmatrix}
    \op\\[3pt]
    \opE
  \end{pmatrix}}=
\begin{pmatrix}
    Z&\Zoffd^{\op\opE}\\[3pt]
    0 & Z^{\opE}
\end{pmatrix}
\begin{pmatrix}
        \op\\[3pt]
        \opE
    \end{pmatrix}\,,\label{eq:opMixingNoGaugeFix}
\end{equation}
with the extra operator 
\bes
  \opE = \frac{1}{g_0^2}\sum\limits_{\mu,\nu,\rho}
  \tr([D_\mu, F_{\mu\nu}]\,[D_\rho, F_{\rho\nu}])\,.
\ees

Since we are just interested in the renormalization matrix $Z$, it suffices to consider only $\R{\op}$, the first block row 
of the above equations. Those define
the renormalized 
$\R{(\CikO)}$, replacing $\op$ with $\R{\op}$.
We write the resulting equations as
\bes
   \label{eq:CikR}
   \R{(\CikO)} = \sum_{j=1}^2 Z_{ij} {\CjkO} + 
   \sum_{l} \Zoffd_{il} {\ClkOred}\,,
\ees
where $\ClkOred$ is formed of the needed redundant operators which mix into $\op$. Without background field, it is the set of $\slashed{\op}$. With background field there is just the operator $\opE$. 
Expanding
\bes
   Z_{ij}&=& \delta_{ij} +  \Zbar_{ij} \frac{\gr^2}{\eps}+
   \ord(\eps^0, \gr^4)\,, \quad
   \Zoffd \;=\;  \Zoffdbar \frac{\gr^2}{\eps} +  \ord(\eps^0, \gr^4)\,,
   \\
   \CikO &=& (\CikO)^{(0)} + \overline{\CikO}\frac{\gr^2}{\eps} +  \ord(\eps^0, \gr^4)\,, \quad
   \Cred \;=\; ({\Cred})^{(0)} +  \ord(\gr^2)\,,
\ees
and requiring the finiteness of $\R{(\CikO)}$, 
the desired $\Zbar_{ij}$ (as well as $\Zoffdbar$) are obtained as the solution of the linear system of equations
(each $i=1,2$ and all $k$ yield an equation),
\bes
   \sum_{j=1}^2 \Zbar_{ij}  (\CjkO)^{(0)} + 
   \sum_{l} \Zoffdbar_{il} ({\ClkOred})^{(0)} =  
   -\overline{\CikO}\,.
\ees

There is one subtlety in applying the above. 
The equations assume that the observables 
$\CjkO$ are infrared finite. 
With the chosen on-shell kinematics in the first case, this is,
however, not true  and the $1/\eps$ terms contain in principle
a mix of ultraviolet and infrared divergences. 
Therefore we use the by now common following trick, called {\it infrared rearrangement}
\cite{Misiak:1994zw,Chetyrkin:1997fm,Luthe:2017ttc}.
For each loop integral, we rewrite 
the denominators in the form
\begin{equation}
 \label{eq:irreaar}
\frac{1}{(k+p)^2}=\frac{1}{k^2+\Omega}-\frac{2kp+p^2-\Omega}{(k^2+\Omega)(k+p)^2}\,,
\end{equation}
where $k$ is the loop momentum and  $\Omega$ is an
arbitrary positive constant. 
The second term on the r.h.s.~is one power less ultraviolet divergent and the first one has
no source of infrared divergence. We can usually 
restrict ourselves to the first one
since we are just interested in the ultraviolet
divergences which determine the renormalization.
If necessary, one can apply the transformation repeatedly. 
While for many integrals this trick is not necessary, we carry it
out in all cases, since all integrals are then brought to the
standard form $\int \rmd^D k\left[k^2 +\Omega\right]^{-n} {k_{\mu_1}\ldots k_{\mu_l}}$
up to the finite and infrared divergent parts which we just drop.
Note that the $Z$-factors 
are independent of $\Omega$. We have used this throughout
as a check on our results. 

The computation was carried out with the help of computer algebra packages. Feynman graphs were 
generated by
\QGRAF/~\cite{Nogueira:1993,Nogueira:2006pq}, formally
treating the operator insertions with the help of 
additional non-propagating scalar fields, $\varphi_i(x)$, called ``anchor'',
through
additional terms $\sum_i\varphi_i(x)\op_i(x)$ in the Lagrangian.
The Feynman rules 
were generated using \FORM/~\cite{Vermaseren:2000nd}, which we also used
for tricks such as  \eq{eq:irreaar},
to reduce the Feynman graphs to standard one-loop integrals,
and to isolate the $1/\eps$ poles.

The computed two-point and three-point functions with operator
insertions are shown schematically in \fig{f:diagrams}.
We checked explicitly that the results for both cases, non-zero $q$ vs. background field, agree. They read
\bes
  \Zbar &=& \frac{\CA}{(4\pi)^2} 
  \begin{pmatrix}
     7/3 & 0 \\
     -7/15 & 21/5 
  \end{pmatrix}\,.
\ees
The element $\bar Z_{11}$ agrees with the value found in the literature~\cite{Narison:1983}.
For completeness we also report the mixing terms~($\CA=\mathrm{N}$ for gauge group SU(N))
\bes  
  \Zoffdbar^{\op\slashed{\op}} &=& 
  \frac{\CA}{(4\pi)^2} \begin{pmatrix}
  -6 & 0 \\
  -21/20 & -9/5 
\end{pmatrix}
\,,\quad
  \Zoffdbar^{\op\opE} = \frac{\CA}{(4\pi)^2} \begin{pmatrix}
  \frac{23}{6}-\frac{3}{2\lambdar} \\
  \frac{7}{15}-\frac{1}{2\lambdar}
\end{pmatrix}
\,,\\
Z^{\opE} &=&  1+ \frac{\CA}{(4\pi)^2} \left(\frac{5}{4}-\frac{3}{4\lambdar} \right)\frac{\gr^2}{\eps}\,.
\ees
We read off that the choice of basis,
\bes
    \base_1= \op_1\,, \quad  \base_2=-\frac14 \op_1+\op_2\,, 
\ees
renormalizes without mixing at one-loop order,
\bes
   \base_i^\mathrm{R}&=&[1+\Zbar^\base_i  \frac{\gr^2}{\eps} ]
   \,\base_i +\rmO(\gr^4)\,,
 \quad  
   \Zbar^\base_1=\frac73\frac1{(4\pi)^2}\,,\quad \Zbar^\base_2=\frac{21}5 \frac1{(4\pi)^2}\,.
\ees
The anomalous dimensions of \eq{eq:gammahat} are\footnote{
At one-loop order we have 
$\gamma_i = 2b_0\hat\gamma_i=\Zbar^\base_i$. }
\bes \label{eq:gammahatYM}
   \hat\gamma_1=7/11\approx 0.636\,,\quad \hat\gamma_2=63/55\approx 1.145 \,,
\ees
independent of the number of colors.
 
\begin{figure}\centering
\includegraphics[width=0.4\textwidth]{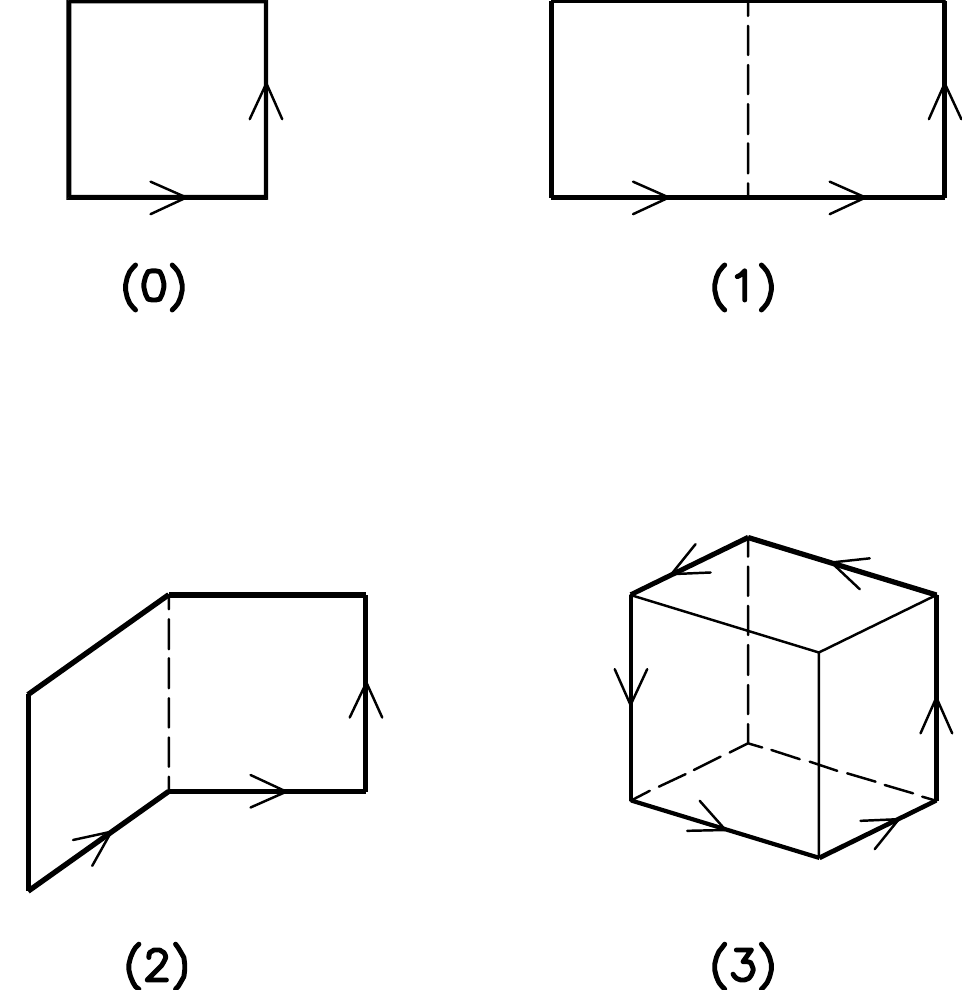}
\caption{Graphical representations~\cite{Weisz:2010nr} of the loop geometries contributing to common lattice gauge actions.}\label{f:gaugeActionTerms}
\end{figure}
\pagebreak
\section{Matching to lattice actions}
\label{s:match}
The final ingredient needed to predict the form of
the cutoff effects are the 
coefficients of the higher dimensional operators in the
effective Lagrangian, step ``\ref{it:match}.'' in \sect{s:sym}. 
At leading order of perturbation theory
considered here, we just need the lowest order coefficients
$\csym_i^{(0)}$ of the functions $\csym_i$, \eq{eq:ciexp}.
At tree-level, no divergences occur in the path integral. One may therefore perform a naive classical
expansion of the lattice action in $a$, 
setting $U(x,\mu) = \rme^{a A_\mu(x)}$ with a smooth continuum gauge field $A_\mu$. This expansion has been carried out 
by L\"uscher and Weisz~\cite{Luscher:1985} for a set of gauge actions, in particular for those
consisting of the lattice loops depicted
in \fig{f:gaugeActionTerms}. For each of these
loops one sums over all lattice points corresponding
to the lower left corners in the graph and over all
orientations on the lattice, e.g. for the plaquette term (0)
one sums over $\mu>\nu$, for the rectangle (1) over $\mu\ne\nu$ 
etc.  There are 6,12,16,48 orientations for the loops (0),(1),(2),(3).
Apart from the overall pre-factor $2/g_0^2$, we denote their coefficients at $g_0 \to 0$ as $e_i,\, i=0,1,2,3$ (in 
Ref.~\cite{Luscher:1985} they are denoted $c_i(0)$). With
\bes
   e_0+8e_1+8e_2+16e_3 = 1\,,
\ees
the leading term in the $a$-expansion,
\bes
   \Slat^\mathrm{class} &=& \int\rmd^4x \left\{ \Lcont(x) + 
   a^2 \sum_{i=1}^2 \omega_i \op_i(x) + \ldots \right\} \,,
\ees
has the conventional normalization. The ellipses summarize  terms that vanish upon the use of the equation of motion
and higher orders in $a$.
From Table~2 of \cite{Luscher:1985} we find 
\\
\bes
    \csym_1^{(0)} &=& \omegasym_1^{(0)}+\frac14 \omegasym_2^{(0)}=\frac1{48} +\frac14 e_1 +\frac13 e_2 
    -\frac14 e_3 \,, 
    \\
     \csym_2^{(0)} &=& \omegasym_2^{(0)} = \frac1{12} + e_1 - e_3 \,.
\ees
 
\begin{table}[h]
  \centering
  \begin{tabular}{cccccc}
    \toprule
    action & $e_1$ & $e_2$  & $e_3$  & $\csym_1^{(0)}$ & $\csym_2^{(0)}$ \\
    \midrule
    Wilson, \eq{eq:Slat} & 0 & 0 & 0 & $\frac1{48}$ & $\frac1{12}$ \\
    Symanzik improved & $-\frac1{12}$ & 0 & 0 & 0 & 0 \\
    Iwasaki \cite{Iwasaki:2011np} & $-0.331$ & 0 & 0 & $-0.0619$ & $-0.2477$\\
    DBW2 \cite{deForcrand:1996bx,Takaishi:1996xj} & $-1.4088$ & 0 & 0 & $-0.3314$ & $-1.3255$\\
\bottomrule
\end{tabular}
  \caption{Commonly used gauge actions and their coefficients of
  the operators $\base_1, \base_2$ in the SymEFT. 
  The row ``Symanzik improved''
  applies to all actions with leading order in $g_0^2$ coefficients as 
  specified there. 
}
 \label{t:ci0}
\end{table}
 
The standard Wilson plaquette action, \eq{eq:Slat},
has $e_0=1$, $e_1=e_2=e_3=0$ and both $\base_1$ and $\base_2$ 
contribute to the order $a^2$. Symanzik improved actions
have $\csym_i^{(0)}=0$ by design. Other actions such as 
the Iwasaki action and the ``DBW2'' action lead to quite 
large coefficients. We show a summary in \tab{t:ci0}.
All considered lattice actions just have the plaquette and the 
rectangle terms. This turns out to lead to 
vanishing coefficients $e_2,e_3$ and in the classical $a^2$ expansion 
only $\op_1$ contributes in the $\op_i$ basis ~\cite{Luscher:1985}.
As discussed before we have to go to the 
basis $\base_i$ with diagonal renormalization at one-loop. 
The relevant coefficients for the asymptotics are then related, $\csym_2^{(0)}=4\csym_1^{(0)}$.

\section{Examples for the asymptotic behavior}
For convenience we combine here the main results
of the previous two sections and discuss some
interesting sample applications.

\subsection{Generic form for spectral quantities}

The cases considered in \tab{t:ci0} are probably the most
relevant for the Yang-Mills theory. Since they all satisfy
$
\csym_2^{(0)}=4\csym_1^{(0)}\,,
$
we have the form
\bes
   \Delta_\obs(a) &=& -a^2 \csym_1^{(0)} \left\{
   \left[\,2b_0 \gbar^2(a^{-1})\right]^{\gammahat_1}  \melrgi_{\obs,1}
   + 4  
   \left[\,2b_0 \gbar^2(a^{-1})\right]^{\gammahat_2}
   \melrgi_{\obs,2}
   \right\} \times
   \nonumber \\ && \times
   \,[1 +\ord(\gbar^2(a^{-1})]\, \text{ for Wilson, Iwasaki, DBW2 actions.}
   \label{eq:DeltaPleadgen}
\ees 
The entire computed leading behavior 
only depends on the coefficient $\csym_1^{(0)}$.
While we cannot predict the relative contribution
of the two powers $\gammahat_{1},\gammahat_2$
because they depend on the non-perturbative matrix 
elements $\melrgi$, their mixture is the same for any 
of the three different actions. The only 
action dependence is in the coefficient of the rectangle
term (geometry (1) of \fig{f:gaugeActionTerms}) and thus the leading cutoff effects have a relative size
\bes
   \text{ Wilson : Iwasaki : DBW2} &\approx& 
    1\;:\;(-3)\;:\;(-16)\;.
\ees

\begin{figure}
\centering
\includegraphics[width=0.495\textwidth]{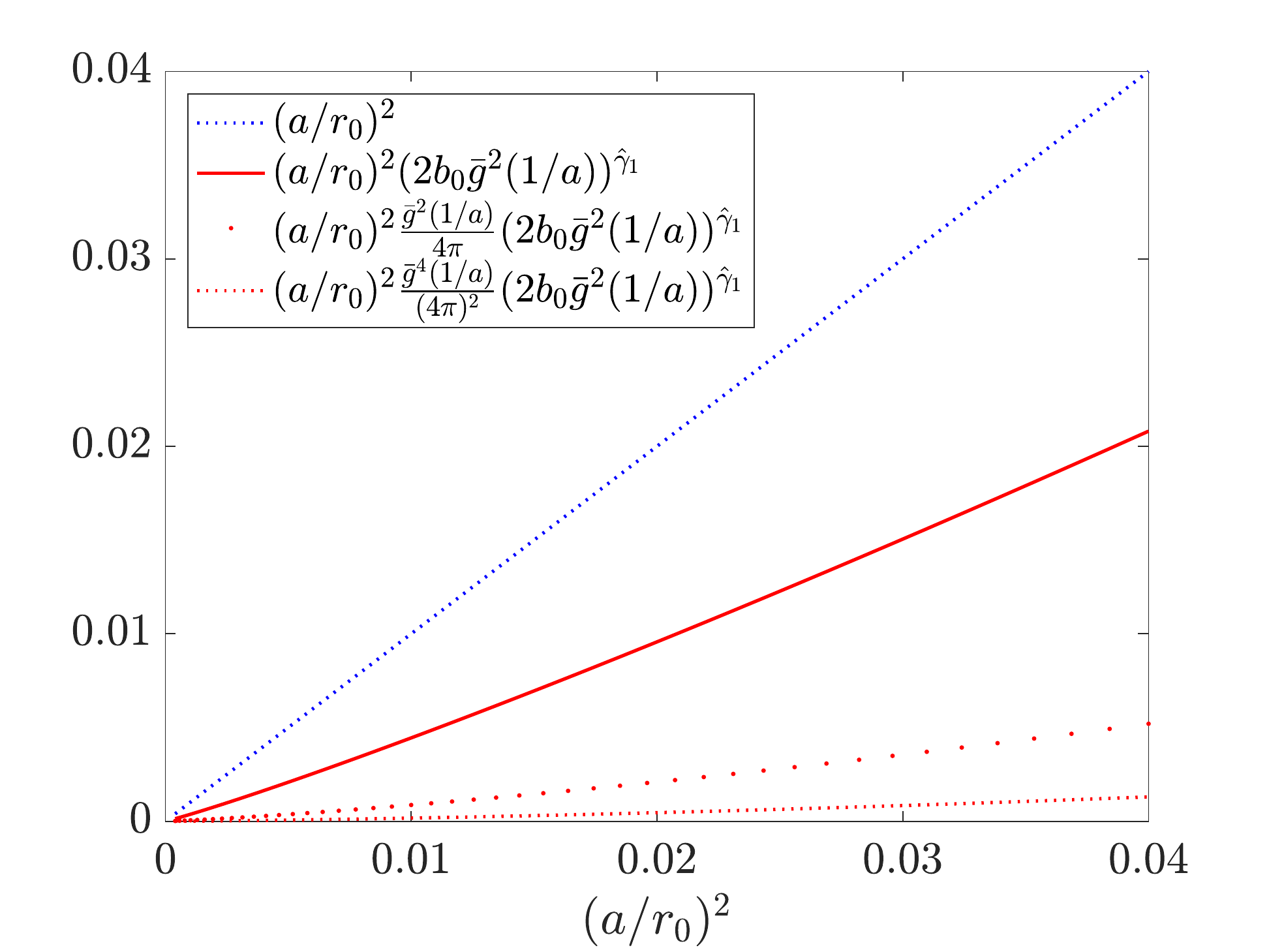}
\includegraphics[width=0.495\textwidth]{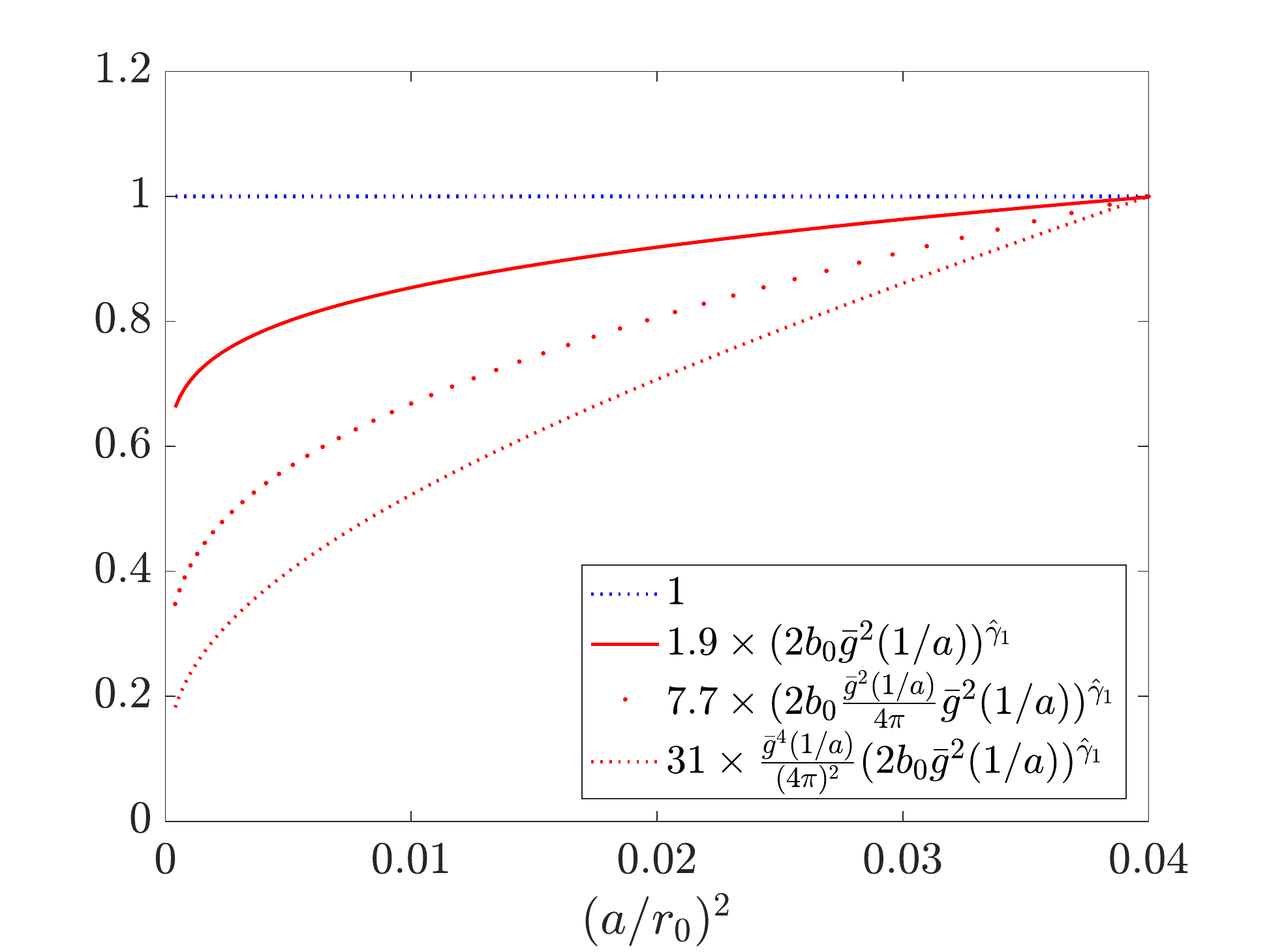}

\caption{Illustration of discretization errors, $\Delta_\obs(a)$,
\eq{eq:DeltaPlead} compared to naive $a^2$ behavior.
We use $\alpha(5/r_0)=\gbar^2(5/r_0)/4\pi=0.25$, 
where $r_0\approx0.5\,\fm$~\cite{Sommer:1993ce} and set matrix elements to one in units of $r_0$ and set $c_1^{(i)}=1$.
On the right, we drop the overall naive power of $a^2/r_0^2$ and normalize at $a/r_0=1/5$ such that the shape is clearly visible. 
}
 \label{f:Delta_lead}
\end{figure}

For a Symanzik improved action, the property
$\csym_2^{(0)}=\csym_1^{(0)}=0$ and additionally for a one-loop 
improved action $\csym_2^{(1)}=\csym_1^{(1)}=0$ 
means 
\bes
   \label{eq:DeltaPlead}
   \Delta_\obs(a) &=& -a^2 \sum_{i} \csym_i^{(n)} \left[\gbar^2(a^{-1})\right]^n 
   \left[\,2b_0 \gbar^2(a^{-1})\right]^{\gammahat_i}  \melrgi_{\obs,i}\,\times
   \\ && \times [1 +\ord(\gbar^2(a^{-1})]\,,
\ees
where $n=1$ for a tree-level improved action
and $n=2$ for a one-loop improved action and $n=0$ 
without perturbative improvement. 
We illustrate the $a$-behavior in 
\fig{f:Delta_lead}. One notices that over a typical range of $a$ from $a=0.1\,\fm$ 
to $a=0.04\,\fm$, one has 20, 40, 60\% 
(for $n=0,1,2$) 
faster than $a^2$ reductions of $\Delta_\obs(a)$
as compared to the naive $a^2$. 

We remind the reader, that gradient flow observables are 
excluded and that we have
restricted ourselves to energy levels.

\subsection{Short distance observables}
Let us now consider the special case of a dimensionless short distance observable 
depending on a single physical length scale $r$.
A simple example is 
$   
   \obs_\mathrm{F} = \frac{4\pi}{C_\mathrm{F}} r^2 F(r)\,,
$
with $F(r)$ the force between static quarks assumed
here to be defined in terms of a discrete derivative
of the potential which is correct up to order $a^4$ 
errors.\footnote{Otherwise, if  $\ord(a^2)$ errors are associated with the definition of the lattice derivative, these can be taken into account explicitly.} 
In particular, we are interested in the region of small 
$r$, 
which has two consequences. The ratio
$a/r$ which determines the discretization errors is
not as small as in the large distance region. 
The discussion of discretization errors is thus 
particularly important. Second, not only the continuum
$\obs(\Lambda r,0)$ can be expanded in 
perturbation theory, but also the quantity at
finite $a/r$ - both in lattice theory 
and in SymEFT. We want to summarize what one can 
learn from this.

The perturbative expansion in the lattice
theory is expected to be of the form \cite{Symanzik:1982dy,Symanzik:1979ph}
\bes
  \Delta_\obs(\Lambda r,a/r) &=&  \obs(\Lambda r,a/r)  - \obs(\Lambda r,0)  
  \nonumber \\
  &=&    \label{eq:DeltaPl}
\obs(\Lambda r,0) \; 
     [ \delta_0(a/r)  
      + \delta_1(a/r) \,\glat^2(r^{-1})  
      +\ldots ]
      \\
      &&\delta_l(a/r) =\frac{a^2}{r^2} \sum_{k=0}^l p_{lk} \log(a/r)^k +\ord((a/r)^4) \,.         
\ees
On the other hand in SymEFT with renormalization group improvement, 
dropping the $\ord(\glat^2(a^{-1}))$ 
corrections, we have
\bes
   \label{eq:DeltaPshortd}
   \Delta_\obs(\Lambda r,a/r) &=& -\frac{a^2}{r^2} \sum_{i} \csym_i^{(0)}  
   \left[\,2b_0 \glat^2(a^{-1})\right]^{\gammahat_i}  [r^2 \melrgi_{\obs,i}(r)]\,\,
   \\ &=& -\frac{a^2}{r^2} \obs(\Lambda r,0)\sum_{i} \csym_i^{(0)}  
   \left[\frac{\glat^2(a^{-1})}{\glat^2(r^{-1})}\right]^{\gammahat_i} \,K_i(r)   
   \,, \quad \\ \nonumber && K_i(r)=\frac{r^2 \mel^\mathrm{R}_{\obs,i}(r;\mu)}{\obs(\Lambda r,0)}\,, \quad \mu=r^{-1}\,,
\ees
where the second argument $\mu$ in $\mel^\mathrm{R}$ is the renormalization
scale of the operator $\base_i^\mathrm{R}$.

For comparison to the fixed order perturbation theory form \eq{eq:DeltaPl} we
expand (remember $\hat\gamma_i=\gamma_i^{(0)}/(2b_0)$)
\bes
   \label{eq:fop}
   \left[\frac{\glat^2(a^{-1})}{\glat^2(r^{-1})}\right]^{\gammahat_i} &=& 1 + \gamma_i^{(0)} \log(a/r)\,\glat^2(r^{-1})\,+\,\ord(\glat^4)\,,
   \\
   K_i(r) &=& 
   [K_i^{(0)} + K_i^{(1)} \glat^2(r^{-1}) + \ord(\glat^4)]\,,
\ees
and find
\bes
    \label{eq:p0}
    p_{00} &=& -\sum_i \csym_i^{(0)}K_i^{(0)} \,,
    \\
    \label{eq:p1}
    p_{10} &=& -\sum_i \csym_i^{(0)}K_i^{(1)} - \sum_i \csym_i^{(1)}K_i^{(0)} 
    \quad p_{11} = -\sum_i \csym_i^{(0)}K_i^{(0)}\gamma_i^{(0)}    
    \,.
\ees

This demonstrates the standard use of EFT in the perturbative
domain. The EFT description and computation is more efficient
since first of all it provides renormalization group improvement (l.h.s. of \eq{eq:fop})
and second even the computation of coefficients $p_{lk}$ may be simplified.
Apart from the one-loop matching coefficients of the action, $\csym_i^{(1)}$,
which can be computed by matching any convenient set of observables, 
only continuum perturbation theory quantities appear on the 
r.h.s. of \eq{eq:p0},~\eq{eq:p1}.

\subsubsection*{Improved observables}\addcontentsline{toc}{subsubsection}{Improved observables}
\label{s:improbs}
For short distance observables it is rather
common to attempt a reduction of lattice 
spacing effects at the level of the expectation values
instead of at the level of the action. For the static potential or $\obs_\mathrm{F}$, we refer the reader to~\cite{Sommer:1993ce,Necco:2001xg}. 
Examples with higher orders in perturbation theory 
and with a combination of improvement of action
and observable are found for example in
\cite{DeDivitiis:1994yz,Bode:1999sm,Alexandrou:2015sea,DallaBrida:2018rfy}. 

To illustrate what is gained by considering SymEFT, it is sufficient to
define a tree-level improved short distance observable,
\bes \label{eq:Pimpr}
  \obs^\mathrm{impr}(\Lambda r,a/r) &=& 
  \frac {\obs(\Lambda r,a/r)}{1+\delta_0(a/r)} 
  =\frac {\obs(\Lambda r,a/r)}{1-\frac{a^2}{r^2}  \sum_i\csym_i^{(0)}K_i^{(0)}} +\ord(a^4/r^4)\,.
\ees
By construction, cutoff effects in fixed 
order perturbation theory are then suppressed by one
power of $\glat^2$ (all orders in $a/r$) and therefore also
the coefficient $p_{00}$ of $a^2/r^2$ vanishes irrespective of the action. 
However, 
this neither means that the leading term ($i=1$) in 
\eq{eq:DeltaPshortd} vanishes nor that the sum
of the two $\ord(a^2)$ terms does. 
The sum of the two terms vanishes only 
for $a=r$, which is not at all where the $a^2$ expansion
is applicable. In fact, inserting the denominator in 
\eq{eq:Pimpr} into 
 \eq{eq:DeltaPshortd} one obtains
\bes
   \label{eq:DeltaPshortdTLI}
   \Delta_{\obs^\mathrm{impr}}(\Lambda r,a/r) &=& 
   - \frac{a^2}{r^2} \,\obs(\Lambda r,0)\,\sum_i
   \left\{ 
   \left[\frac{\glat^2(a^{-1})}{\glat^2(r^{-1})}\right]^{\gammahat_i} -1
  \right\}
   \, K_i^{(0)}\csym_i^{(0)} \, \,.
\ees
The effect of tree level improvement is the  
subtraction of the $1$ in the curly bracket. 
For intermediate $a/r$, this will reduce the magnitude 
(and change the sign) of each term in the sum over $i$.
However, asymptotically, for very small $a/r$, the tree level improvement leads to an increase of the $a^2$ effects.
This behavior is  tied to the sign of 
the $\gammahat_i$. For negative $\gammahat_i$, we would always have a reduction of the magnitude of the terms.

Usually the terms $K_i^{(0)}\csym_i^{(0)}$ are known
individually and one can divide out the complete leading order term, 
\bes
  \label{eq:PRGimpr}
  \obs^\mathrm{RG-impr}&=& 
  \frac{\obs}{1- \frac{a^2}{r^2}\,\sum_i 
   \left[\frac{\glat^2(a^{-1})}{\glat^2(r^{-1})}\right]^{\gammahat_i}    \, K_i^{(0)}\csym_i^{(0)}  } \,,
\ees
and
have a renormalisation group and tree level improved observable.  
It then has leading corrections which are truly of order 
$ \Delta_\obs/\obs \sim \frac{a^2}{r^2} \glat^2(r^{-1})\left[\frac{\glat^2(a^{-1})}{\glat^2(r^{-1})}\right]^{\gammahat_1} $
as the name tree level improvement suggests.

\begin{figure}
\centering
\includegraphics[width=0.695\textwidth]{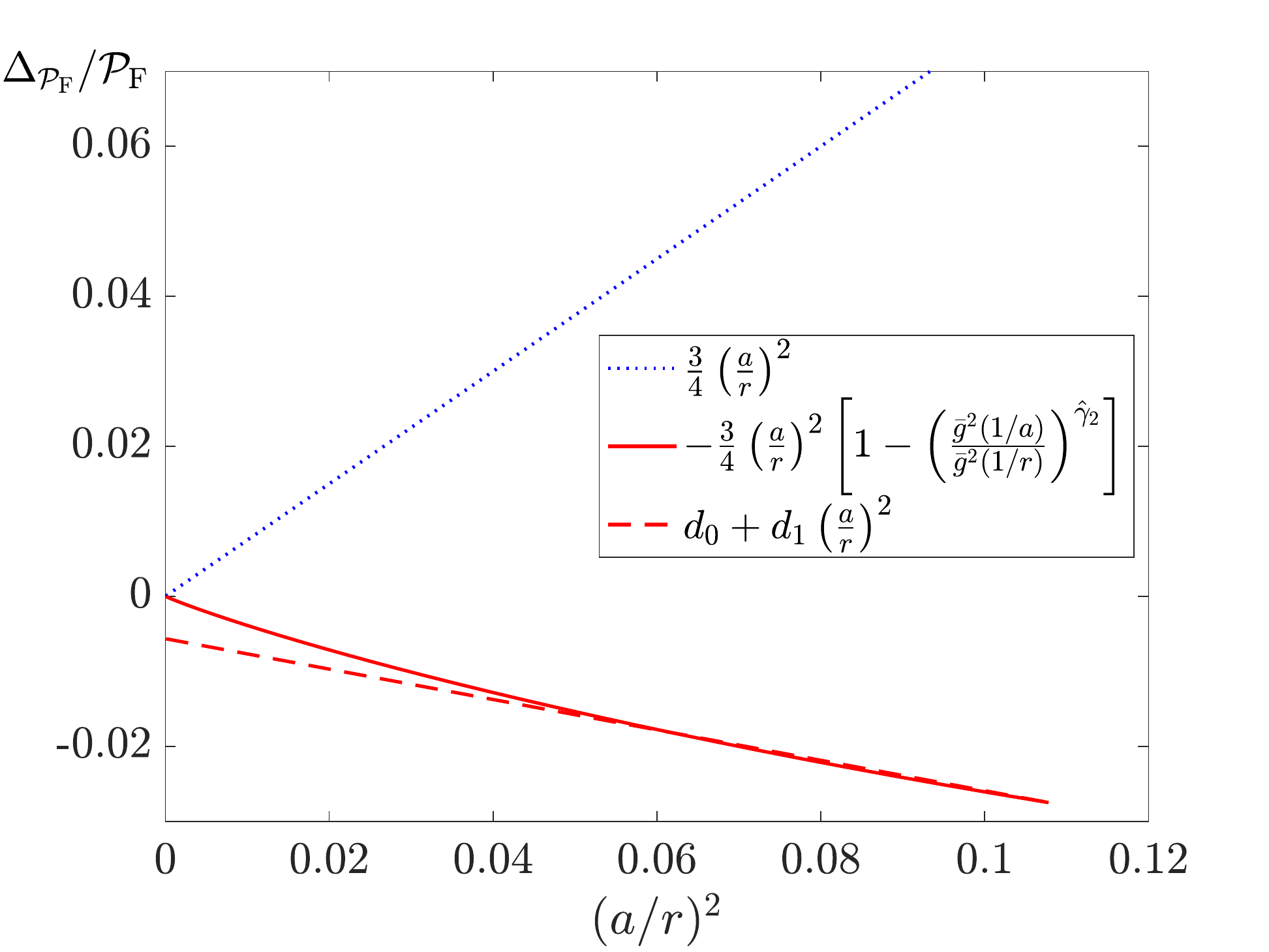}
\caption{Leading order discretization errors, $\Delta_{\obs_\mathrm{F}}(a)/\obs_\mathrm{F}$ of the 
static force,
see the text.
We use $\alpha(1/r)=\glat^2(1/r)/(4\pi)=0.2$, 
and $K_1^{(0)}=0$, $K_2^{(0)}\csym_2^{(0)} = -3/4$,
corresponding to the Wilson plaquette action and the force along
a lattice axes.
The dotted line represents fixed order perturbation theory, the full line the remainder (on top of fixed order) predicted by SymEFT, and the dashed line shows
a rough approximation, linear in $a^2$,  to that latter.}
 \label{f:Delta_lead_force}
\end{figure}

We return to $\obs_\mathrm{F}$.
In this special case, the O(4) invariant operator $\op_1=\base_1$ 
does not contribute at tree level,
$K_1^{(0)}=0$.  Specializing to the Wilson plaquette
action and the force along a lattice axes,
we have $\csym_2^{(0)}=1/12$ and $ K_2^{(0)}=-9$. 
If one chooses a different direction on the lattice,
e.g. a body-diagonal, 
the matrix element $K_2^{(0)}$ is smaller, 
but the finite difference defining the force on
the lattice has a larger discretization length. 
The various terms are illustrated in \fig{f:Delta_lead_force}. 
The dotted line is the  fixed order 
perturbation theory for $\Delta_{\obs_\mathrm{F}}/\obs_\mathrm{F}$
 and the full curve the remainder, \eq{eq:DeltaPshortdTLI}. 
 The dashed line
 shows a rough linear approximation to the latter at larger $a$. It extrapolates to a 
 small value  of $-0.6\%$ at $a=0$. We may think of 
 this as an example for the relative error one makes
 by approximating the cutoff effects 
 of the tree-level improved observable linear in $a^2$.\footnote{Usually the tree-level improved force is defined through an improved distance \cite{Sommer:1993ce} $r_\mathrm{I}$.
 At the level of $a^2$ terms this is equivalent to \eq{eq:Pimpr}.} 
  Interpreting $\obs_\mathrm{F}$ as a running coupling   as explained for example in 
 \cite{Necco:2001gh}, this intercept represents 
 a systematic (relative) uncertainty on the 
 coupling. It  translates into an about 1.5\% 
 error in the $\Lambda$-parameter of the theory,
 which is not entirely irrelevant given today's precision 
 of results for it. Needless to say that the full logarithmic term
 \eq{eq:DeltaPshortdTLI} is better eliminated by use
 of \eq{eq:PRGimpr}.

\section{Schr\"odinger functional}
\label{s:SF}

\newcommand{\cbt}{\tilde c_\mathrm{b}}
\newcommand{\cb}{\omega_\mathrm{b}}
\newcommand{\csymb}{\csym_\mathrm{b}}
\newcommand{\cbtzero}{\cbt^{(0)}}
\newcommand{\cbtone}{\cbt^{(1)}}
\newcommand{\cbttwo}{\cbt^{(2)}}
\newcommand{\cbzero}{\cb^{(0)}}
\newcommand{\cbone}{\cb^{(1)}}
\newcommand{\cbtwo}{\cb^{(2)}}
\newcommand{\ctzero}{\ct^{(0)}}
\newcommand{\ctone}{\ct^{(1)}}
\newcommand{\cttwo}{\ct^{(2)}}
\newcommand{\af}{a_\mathrm{f}}

Short distance observables of particular interest 
can be defined in the  
Schr\"odinger functional \cite{Luscher:1992an}.
Fixed order perturbation theory 
has been used extensively to study
discretization errors in this environment.
Here we consider their renormalization group
improvement through SymEFT. We just consider 
the pure gauge theory and the Schr\"odinger functional\ with an abelian background field, where - as we will see - we do not have to deal with operator mixing.

In the lattice regularization, the Schr\"odinger functional\ can be defined by the 
path integral with the action, 
\bes \label{eq:SlatSF}
  S^\mathrm{SF}_\mathrm{lattice} &=& \frac2{g_0^2}\,    \sum_{0\leq x_0\leq T-a} 
  \sum_\vecx \sum_{\mu>\nu=0}^3\,   
   p(x,\mu,\nu)
   \\ && + a\, (\ct(g_0)-1) \,a^3 \sum_\vecx  \, [ 
   \op_\mathrm{b}^\mathrm{l}(0,\vecx) + \op_\mathrm{b}^\mathrm{l}(T-a,\vecx)] \,, 
\ees
with 
\bes
   \op_\mathrm{b}^\mathrm{l}(x_0,\vecx)
   = \frac2{g_0^2}\,\frac1{a^4}\, \sum_{k=1}^3 p(x,k,0) \,.
\ees
Space-time is a cylinder in the sense that we have 
periodic boundary conditions in space with period $L$ 
and Dirichlet boundary conditions on the time-slices $x_0=0$
and $x_0=T$,
\bes
   \left. U(x,k)\right|_{x_0=0} = \rme^{a C_k(L,\eta)}\,,
   \quad  \left. U(x,k)\right|_{x_0=T} = \rme^{a C'_k(L,\eta)}\,.
\ees
For details we refer to \cite{Luscher:1992an}, but we note 
that the dimensionless $L C_k(L,\eta)$ is just a function 
of the dimensionless parameter $\eta$ (and a here irrelevant second parameter $\nu$) 
and that the field strength $F_{kl}$ vanishes at the two boundaries.

Under these conditions, which have been imposed for all numerical
applications so far, 
the SymEFT for the Yang-Mills Schr\"odinger functional
is given by the formal continuum action
\bes \label{eq:SSymSF}
   \Seff^\mathrm{SF} &=& \int\rmd^3 \vecx \left\{ \int_0^T\rmd x_0 \Lcont(x) + 
   a\, \cb\, [ 
   \op_\mathrm{b}(0,\vecx) + \op_\mathrm{b}(T,\vecx)]  \right\} +\ord(a^2) 
\ees
with 
\bes
\label{eq:Bb}
  \op_\mathrm{b}(x) &=& -\frac{1}{g_0^2}\tr(F_{0k}(x) F_{0k}(x))\,. 
\ees
The presence of the boundary terms
in \eq{eq:SSymSF} is the reason for including the 
corresponding extra term proportional to $\ct$  
in the lattice formulation:
the coefficients $\ct^{(i)}$ in
\bes 
  \label{eq:ctterm}
   \ct(g_0) =\ctzero  + \ctone g_0^2 + \ord{(g_0^4)} \,,
\ees
can be chosen such that $\cb$ vanishes and there
are no linear terms in $a$ in the lattice Schr\"odinger functional  at 
the corresponding order in perturbation theory~\cite{Luscher:1992an}.

A prominent observable in the 
Schr\"odinger functional is the running coupling 
\bes
   \gbar^{-2}(L^{-1} ) = \frac1{k} \langle S' \rangle \,,\;\text{ with }
   S' = \left. \frac{\partial S}{ \partial \eta}\right|_{\eta=0} \,.
\ees
with $k$ such that $\gbar^2 = g_0^2+\ord(g_0^4)$.
We want to discuss its $a$-effects as an example. The definition 
of the $a$-effects requires to first 
renormalize. We here do this by lattice minimal subtraction,
\bes 
 \label{eq:glat}
 \glat^2(\mu) &=& Z_\mathrm{g}(\glat,a\mu) g_0^2\,,
 \quad Z_\mathrm{g}(\glat,a\mu) = 1 -2b_0  \log(a\mu)\glat^2(\mu)  
+\ord(g^4)\,.
\ees
We can then define the function
\bes
    K(\glat^2(\frac1L ), \frac{a}L ) = \gbar^{-2} \,, 
\ees
which relates the renormalized couplings of the two schemes. 
It has a continuum limit and discretization errors
\bes     
    \Delta K(\glat^2,\frac{a}L )= K(\glat^2,\frac{a}L ) -K(\glat^2, 0) \,.
\ees    
They have the expansion 
\bes
   \frac{\Delta K(\glat^2,\frac{a}L )}{K(\glat^2,0)} =\frac{a}{L}  
   \,[p_{00} + (p_{10} + p_{11}\log(\frac{a}L ) )\glat^2(\frac1L ) +\ord(g^4) ] + \ord((a/L)^2)\,,
\ees
where analogously to before SymEFT predicts 
\bes
   p_{11} =  \gamma^{(0)}_\mathrm{b} \, p_{00} \,.
\ees
An explicit one-loop computation showed that~\cite{Luscher:1993gh}
\bes 
   p_{00}&=&2\,(\ctzero-1)\,,
   \\
   p_{10}&=& 2 \times (\ctone + 0.0890(2))  \,, \text{ for } \ctzero=1\,.
\ees
Thus $\ctzero=1,\, \ctone = - 0.0890(2)$ 
leads to the absence of linear $a$-effects at one-loop. For this reason 
the perturbative computations have been carried out with
$\ctzero = 1$ and from the published one-loop computation 
we do not have access to  $\gamma^{(0)}_\mathrm{b}$.

As done in \sect{s:AD}, the standard way to compute $\gamma^{(0)}_\mathrm{b}$ is to compute
the one-loop renormalization of $\op_\mathrm{b}$. 
Here we extract it indirectly from the results of the 
two-loop computation of ~\cite{Bode:1998hd,BodeThesis}. In contrast to \sect{s:AD}
the computation thus relies entirely on 
the lattice regularization. 
Consider \eq{eq:SlatSF} with a lattice spacing 
$a\to \af$ and then replace
\bes 
    \ct(g_0)-1 \to \zeta \,.
\ees
In this way $\zeta$ acts as a source for the lattice regularized operator 
$\op_\mathrm{b}$. The continuum function 
$
K(\glat^2,0)
$
is given by
\bes \label{eq:A}
     K(\glat^2,0)  =
       \lim_{\af\to 0} \left[ 
     \langle S' \rangle_{\af}\right]_{\zeta=0} 
\ees
and the first order correction in $a$ by
\bes \label{eq:Aa}
     \Delta K(\glat^2,\frac{a}L)=
      a\,\lim_{\af\to 0} 
     \, \left[ \frac1\af \frac{\partial}{\partial \zeta}
     \langle S' \rangle_{\af}^\mathrm{R}\right]_{\zeta=0}
    + \ord((a/L)^2)\,.
\ees
The right hand side of \eq{eq:Aa} is the SymEFT prediction written as 
the continuum limit of the lattice regularized theory
(with spacing $\af$ to distinguish it from $a$). 
Renormalization is indicated by the superscript $\mathrm{R}$.
In addition to \eq{eq:glat}  it affects  
the boundary operator $\op_\mathrm{b}$,
\bes 
 \op_\mathrm{b}^\mathrm{lat} &=& Z_\mathrm{b}(\glat,\af\mu) 
  \op_\mathrm{b}\,, \quad Z_\mathrm{b}(\glat,\af\mu) =
  1-\gamma^{(0)}_\mathrm{b} \log(\af\mu) \glat^2(\mu)  
+\ldots\,.
\ees
We are now ready to extract $\gamma^{(0)}_\mathrm{b}$ from 
the two-loop expansion,
\bes
   \gbar^{-2} &=& g_0^{-2} \, [1 + k_1 g_0^2 + k_2 g_0^4 + \ord(g_0^6)]
   \\
   k_1 &=& - m_1^a + \ctone \frac{2\af}{L} \,,\quad 
   \\
   k_2 &=& - m_2^a - \ctone m_2^b - (\ctone)^2  m_2^c  - \cttwo m_2^d \,,\quad 
\ees
derived in~\cite{Bode:1998hd,BodeThesis} for $\ctzero=1$. We use the asymptotic expansion of the coefficients $m_i^k$ in powers of $\frac{\af}L $ and $\log(\frac{\af}L )$
given in Ref.~\cite{Bode:1998hd,BodeThesis}.
But first we note that with $\langle S' \rangle = k/\gbar^{2}$
we have 
\bes \label{eq:gsq1lp}
   \frac1{\af }\left[  \frac{\partial}{\partial\zeta}
     \langle S' \rangle_{\af}\right]_{\zeta=0}^\mathrm{R}
     &=& \frac1\af Z_\mathrm{b}(\glat,\af\mu) 
      \left[  \frac{\partial}{\partial \zeta}
     \langle S' \rangle_{\af}\right]_{\zeta=0}
     \\ \nonumber
     &=& Z_\mathrm{b}(\glat,\af\mu) \frac k{g_0^2}[\frac2L - \frac1\af  m_2^b(\frac{\af}L) \,g_0^2 +\ord(g_0^4) ] 
     \\
     &=& \nonumber
     \frac k{\glat^2(\mu)}\left[\frac2L -\frac2L \, (\gamma^{(0)}_\mathrm{b}+2b_0) \log(\af\mu) \glat^2(\mu) - \frac1\af  m_2^b(\frac{\af}L) \,\glat^2(\mu) \right]
     \\&& +\ord(\glat^2) \nonumber
\ees
since the computation~\cite{Bode:1998hd,BodeThesis} corresponds 
to $\zeta= \ctone g_0^2 +\ord(g_0^4)$. Finally,
requiring finiteness of \eq{eq:gsq1lp} after inserting
\bes
   \frac1\af  m_2^b(\af/L)  &=& \frac 1L [r_2^b + s_2^b \log(L/\af) + \ord(\af/L) ]\,,
\ees
with~\cite{BodeThesis} $r_2^b=0.1683(8)\,,\; s_2^b = 0.2785(4)$
we obtain $\gamma^{(0)}_\mathrm{b} = s_2^b /2 -2b_0$ and 
\bes \label{eq:gammabres}
     \hat\gamma_\mathrm{b} = 0.000(2) \,.
\ees
Note that this is the anomalous dimension of a boundary 
operator.
Assuming that $\hat\gamma_\mathrm{b} = 0$, exactly,
\Eq{eq:Aa} can now be written in the form (see also \eq{eq:DeltaPshortdTLI}) 
\bes \label{eq:DKasy}
    \Delta K
    &=&
     \frac{a}L [\gbar^2(a^{-1})]^{n_I+1}\,2\,\csymb^{(n_I+1)}\,[1 +\ord(g^2)]   \,, 
\ees
where $\csymb^{(n_I+1)} = -\ct^{(n_I+1)}$ is the leading coefficient
in 
\bes
   \csymb = \csymb^{(n_I+1)}[\gbar^2(a^{-1})]^{n_I+1}
     + \ord([\gbar^2(a^{-1})]^{n_I+2})\,, \label{eq:cbexp}
\ees
namely we are considering a theory where $\ct$ is chosen 
to achieve $\ord(a)$ improvement in perturbation theory, up to and including the 
terms $g_0^{2n_I}$. 
The $\ord(\gbar^2(\frac1L ))$ term in the SymEFT matrix element
is given by $r_2^b \gbar^2/2$, but it comes together with the two-loop anomalous dimension of the boundary operator and
the next order correction in \eq{eq:cbexp}.
Since these are presently unknown, we only show the leading order in $g^2$ in \eq{eq:DKasy}.

In order to compute the non-perturbative 
running of the coupling, one considers the step scaling
function,
\bes
   \Sigma(u,\frac{a}L ) = \left.\gbar^2(1/(2L))\right|_{\gbar^2(1/L)=u}\,,
\ees
where the choice of intermediate renormalization scheme (we chose ``lat'') disappears. 
Its leading discretization errors are (see also \cite{DallaBrida:2018rfy}, App. A)
\bes 
\Delta \Sigma(u,\frac{a}L ) &=& \Sigma(u,\frac{a}L )-\Sigma(u,0)
\\
&=& u \frac{a}{L} \csymb^{(n_I+1)}[\gbar^2(a^{-1})]^{n_I+1}[1 +\ord(u)] 
\label{eq:sffinal}
\ees
Since we have seen that the one-loop anomalous dimension of $\op_\mathrm{b}$ vanishes, this is equivalent to the form
used by the ALPHA collaboration recently \cite{Bruno:2017gxd,DallaBrida:2018rfy}.

\section{Wilson-QCD}
\label{s:Wils}
Let us now briefly discuss the case of the 
original Wilson action for QCD including the Wilson term 
in the fermion action~\cite{Wilson:1974}. While this action is 
hardly used any more in the original form it is still of interest
because there are results in the literature. More importantly,
some large scale computations use the $\ord(a)$-improved version 
with an approximate coefficient of the clover improvement term. 
One can gain information on the scaling of $\delta\obs^\L$, \eq{eq:S2matel}, in that case.

The Wilson quark action breaks chiral symmetry and thus allows
for the dimension five 
Sheikholeslami-Wohlert term\cite{Sheikholeslami:1985ij}

\bes
    \dlatt[1]{\L}(x)= - \omegaswsym \frac18 \,\psibar(x)[\gamma_\mu,\gamma_\nu] F_{\mu\nu}(x)\psi(x) 
\ees
in the SymEFT, \eq{eq:effLagrangian}. In principle there are additional terms proportional to quark masses, but these ``only'' affect
quark-mass dependences \cite{Luscher:1996sc} and are absent when
one takes the continuum limit along a physical scaling trajectory 
defined by, for example, fixed ratios of $\nf$ pseudo-scalar masses in the
$\nf$-flavour theory. 
We here neglect those $\ord(a\mq)$ effects; we set the quark masses to zero.
There are no operators which violate rotational symmetry.
Therefore, there is no mixing at $\ord(a)$ at all.
The prediction for the asymptotic $a$ dependence can then
immediately be written down,
\bes
   \Delta_\obs(a) =  -a  \, \cswsym^{(0)} \left[\,2b_0 \gbar^2(a^{-1})\right]^{\gammahat^\mathrm{sw}} \melrgi \times [1 + \ord(\gbar^2(a^{-1}))]
    \sim  a   \left[\frac{1}{-\log(a\Lambda)}\right]^{\gammahat^\mathrm{sw}} \,. 
\ees
For the standard Wilson action, 
we have $\cswsym = \cswsym^{(0)} +\ord(g^2)$ with $\cswsym^{(0)}=-1$. 
As in \eq{eq:sffinal}, there are additional powers of $\gbar^2(a^{-1})$ when the theory is perturbatively $\ord(a)$ improved \cite{Sheikholeslami:1985ij,Wohlert:1987rf,Luscher:1996sc,Aoki:2003sj}.
We find~\cite{H:inprep} 
($C_\mathrm{A}=\mathrm{N},\; C_\mathrm{F}=(\mathrm{N}^2-1)/(2\mathrm{N})$), 
\bes
   \gammahat^\mathrm{sw} = \frac{15C_\mathrm{F}-6C_\mathrm{A}}{11C_\mathrm{A}-2\nf}
\ees
for the anomalous dimension. It is rather small. For $\mathrm{N}=3$ this is in agreement with~\cite{Narison:1983}.

For the considered case of Wilson fermions, one may also easily 
discuss the relevant contributions from corrections to 
the vector and axial vector, non-singlet, flavor currents. In SymEFT,
they are represented by~\cite{Luscher:1996sc}
\bes
 V^{r,s}_\mu(x)&=&\psibar_r(x) \gamma_\mu \psi_s(x)
  + a \,\omegavsym \,\partial_\nu T^{r,s}_{\mu\nu}(x) \, ,
 \\
 A^{r,s}_\mu(x)&=&\psibar_r(x) \gamma_\mu \gamma_5 \psi_s(x)
 + a \, \omegaasym \,\partial_\mu P^{r,s}(x)\,.
\ees
Matrix elements of interest of the corresponding lattice currents
are, e.g., leptonic decay constants and semi-leptonic 
form factors. Using the anomalous dimensions of the non-singlet pseudo scalar density and the tensor current \cite{Larin:1993tq,Broadhurst:1994se}, their lattice artifacts receive contributions
\bes
   \Delta_\obs^\mathrm{V}(a) &=& a \, \gbar^2(a^{-1})\left[\,2b_0 \gbar^2(a^{-1})\right]^{\gammahat^\mathrm{T}} \melrgi_\mathrm{T} \,[\cvSym^{(1)} +\ord(\gbar^2(a^{-1})]\,, \quad \gammahat^\mathrm{T}  
   =\frac{3C_\mathrm{F}}{11C_\mathrm{A}-2\nf}\,,
   \nonumber \\ \label{eq:DeltaVA}\\[-2ex]
   \nonumber
   \Delta_\obs^\mathrm{A}(a) &=& a \, \gbar^2(a^{-1})\left[\,2b_0 \gbar^2(a^{-1})\right]^{\gammahat^\mathrm{P}} \melrgi_\mathrm{P} \,[\caSym^{(1)} +\ord(\gbar^2(a^{-1})]\,,\quad \gammahat^\mathrm{P} = 
   -3\, \gammahat^\mathrm{T}\,.  
   \ees 
where $\melrgi_\mathrm{T}$ is the RGI matrix element of 
$\partial_\nu T^{r,s}_{\mu\nu}$ of the considered transition and 
$\melrgi_\mathrm{P}$ the RGI matrix element of $\partial_\mu P$. 
There is an extra factor $\gbar^2$, as compared to previous 
expressions, since the $\ord(a)$ term in the classical expansion of the currents vanishes.  The $\omega_{\rm V/A}^{(1)}$ factors are the one-loop matching 
coefficients between SymEFT and the considered lattice theory.
An extended list of references with results for 
improvement coefficients $c_\mathrm{V/A}^{(1)}$ 
for various actions is given in Table 1 of 
\cite{Sommer:2006sj}. 
The case of unimproved lattice currents, e.g. $ V^{r,s}_{\mu,\mathrm{latt}}(x)=\psibar_r(x) \gamma_\mu \psi_s(x)$, can be obtained by setting
$\csym
_\mathrm{V/A}^{(1)}=-c_\mathrm{V/A}^{(1)}$ in \eq{eq:DeltaVA}.
These coefficients are rather small.

\section{Summary}

We have investigated the form of the leading discretization
errors in lattice gauge theory in a few specific cases.
 The 
starting point is the leading contribution to the 
Symanzik effective Lagrangian in the form
\bes 
  \label{eq:effLagrangian2}
\Leff(x)=\L(x)+a^{\nmin}\sum_i \csym_i^{(n_i)}g^{2n_i}\base_i(x)+\ldots\,,
\quad \nmin \geq 1\,,\quad n_i\geq0\,,
\ees 
where the ellipsis denotes higher powers in $g^2$ for each term $i$ as well as  higher powers in $a$.
The basis operators are chosen such that they do not mix at one-loop order and have one-loop anomalous dimensions 
$\gamma_i^{(0)}g^2, \; \gamma_1^{(0)} \leq \gamma_2^{(0)} \leq \ldots$.
Once $\nmin,\, c_i,\, n_i,\, \gamma_i^{(0)}$ are known, the leading correction to the continuum limit of spectral quantities
is
\bes
\label{eq:DeltaPconcl}
  \Delta_\obs(a)   &=& 
  a^\nmin \left[\gbar^2(a^{-1})\right]^{n_1} 
   \left[\,2b_0 \gbar^2(a^{-1})\right]^{\gammahat_1} \, \csym_1^{(n_1)}
   \melrgi_{\obs,1}\,\,
    [1 +\ord([\gbar^{2}(a^{-1})]^{\Delta\gammahat},\gbar^2(a^{-1}))]\nonumber \\ 
     && +\rmO(a^{\nmin+1})\,,
\ees
with $\gammahat_i=\gamma_i^{(0)}/(2b_0)\,,\; \Delta\gammahat=\gammahat_2-\gammahat_1$. The only unknown is 
the $a$-independent 
renormalization group invariant matrix element 
$\melrgi_{\obs,1}$ of the operator $\base_1$. 
The most important ingredient in the formula is the leading 
$\gammahat_1$. In almost all considered cases, we find that 
$\gammahat_1\geq 0$ in stark contrast to the case of the 2d O(3) model
\cite{Balog:2009yj}. This is good news, as the leading corrections accelerate the approach to the
continuum limit compared to the naive classical argumentation
which neglects the overall $\left[\gbar^2(a^{-1})\right]^{n+\gammahat_1}$ factor.

Let us briefly summarize the results for the individual cases considered. 

\begin{itemize}
\item Yang-Mills theory.\\
Discretization effects of order $a^2$ are due to two operators. Their anomalous dimensions, $\gammahat_i$, computed in \sect{s:AD}, are of order one, see \eq{eq:gammahatYM}. In eqs. (\ref{eq:effLagrangian2}- \ref{eq:DeltaPconcl}), the original Wilson action, tree-level and
one-loop Symanzik improved actions have $n_i=0,1,2$
respectively. 
\item Yang-Mills theory with a boundary: Schr\"odinger functional.\\
As discussed in \sect{s:SF} there are linear in $a$ 
discretization errors due to one boundary operator. Using the literature on  perturbation theory for the Schr\"odinger functional, we extracted its anomalous dimension and 
found that it vanishes within uncertainties, $\gammahat_\mathrm{b}=0.000(2)$. 
This means that the fixed order perturbation theory analysis
of discretization errors carried out by the ALPHA collaboration \cite{Bruno:2017gxd} receives no log-corrections at leading order.
\item Wilson $\ord(a)$ effects due to the fermion action.
\\
Here our analysis concerns $\ord(a)$ effects which 
come from an action with perturbative improvement,
i.e. an improvement coefficient $\csw$ determined at
$n$-loop perturbation theory. The Pauli term, found to
be the only contributing operator
by Sheikholeslami and Wohlert, has $n_1=n+1$ in
\eq{eq:DeltaPconcl}. Its anomalous dimension, 
 $\gammahat_1=\gammahat^\mathrm{sw} = \frac{15C_\mathrm{F}-6C_\mathrm{A}}{11C_\mathrm{A}-2\nf} $, 
 could be taken from the literature~\cite{Narison:1983}. It is rather small. Interestingly, as one approaches the conformal window \cite{Nogradi:2016qek} by increasing $\nf$, the anomalous dimension $\gammahat^\mathrm{sw}$ grows. 
 \item Wilson $\ord(a)$ effects due to the flavor currents.
\\
Weak decay (and other) matrix elements receive {\em additional}
discretization errors from correction
terms in the effective weak Hamiltonian. We just considered
the flavor currents with perturbative
$\rmO(a)$ improvement in \sect{s:Wils}. For the 
axial current, the (derivative of the) pseudo-scalar field 
governs the correction term. Its $\gammahat_\mathrm{P}$ is negative, but relatively 
small in magnitude. Since the coefficient of the correction operator
starts at order $g^2$
in perturbation theory, the total logarithmic
modification, $\left[\gbar^2(a^{-1})\right]^n\left[\,2b_0 \gbar^2(a^{-1})\right]^{\gammahat_\mathrm{P}}$, again accelerates convergence due to
$n\geq 1$ and $n+\gammahat_\mathrm{P}>0$. 
For the vector current the $\rmO(a)$ correction involves 
the tensor current
with $\gammahat_\mathrm{T}$ which is positive and
 rather small. This leads to an even better $a$-dependence. Note that this analysis holds also for a 
non-perturbatively improved action but only perturbatively improved currents.
\end{itemize}  

Short distance observables $\obs(r\Lambda)$ with $r\Lambda\ll 1$ are special. Their matrix
elements
$\melrgi_{\obs,i}(r\Lambda)$ are computable 
in renormalized perturbation theory in terms of the coupling at scale $\mu=1/r$ and one can make 
parameter free predictions for the leading corrections. 
As discussed in \sect{s:improbs} the usual tree-level {\em improved observables} do not always lead to a reduction 
of the asymptotic cutoff effects, but this is easy to rectify
such that cutoff-effects are suppressed by one power of
$\gbar^2(r^{-1})$ at short distances.

As a broad conclusion, our results are very positive because 
the so-far known logarithmic corrections are relatively weak. 
This lends support to some of the continuum extrapolations 
performed in the literature. 
For example, the BMW collaboration has performed
continuum extrapolations of data obtained with tree-level 
coefficient, $\csw=1$  of the Sheikholeslami-Wohlert term~\cite{Durr:2010vn}.
In principle, the asymptotic behavior is then
$\csym_\mathrm{sw}^{(1)} \gbar^2(a^{-1})\left[\,2b_0 \gbar^2(a^{-1})\right]^{\gammahat^\mathrm{sw}}$. In one of their continuum 
extrapolations they used this form but with $\gammahat^\mathrm{sw}\to 0$, which we now see is a rather good approximation.
Of course, the difficult question in such extrapolations 
is whether one is in the region where the asymptotics dominates. 
For this reason they also used alternative extrapolation functions.

Despite the small values of $\gammahat$ that we found, 
with tree-level or one-loop Symanzik
improved action, the $\left[\gbar^2(a^{-1})\right]^n\left[\,2b_0 \gbar^2(a^{-1})\right]^{\gammahat_1}$ effects are non-negligible 
when MC results are precise, see the right part of \fig{f:Delta_lead}. In any case, when the leading behavior 
is known, it should be incorporated into the fit function.
Still, we emphasize that the asymptotically leading
behavior can be predicted, not the region where exactly this
dominates over formally suppressed terms.

Of course the most interesting application of SymEFT is lattice 
QCD with 
$\nmin=2$ in \eq{eq:logcorr}. In that case the basis of contributing operators
is considerably larger. Work on determining their
anomalous dimensions is in progress~\cite{H:inprep}. 
Also Gradient flow observables are of high interest. Their discretization
errors are surprisingly large~\cite{Sommer:2014mea,Ramos:2014kka,DallaBrida:2016kgh}. Now that it is known that 
standard pure gauge theory operators are not the source 
of this behavior, since they have positive $\gammahat_i$, a
natural suspicion is that there is  
an unusually large and negative anomalous dimension $\gammahat$ of the additional dimension six
operator at $t\to0$, present in the 5-d formulation 
of the Gradient Flow, 
see \cite{Ramos:2014kka} for more details. We also plan to investigate this issue.

\textbf{Acknowledgements.} We thank Hubert Simma and Kay Sch\"onwald for many discussions.

\vskip 0.3cm

\noindent

\providecommand{\href}[2]{#2}\begingroup\raggedright\endgroup


\end{document}